\documentclass[11pt]{article}
\pdfoutput=1
\usepackage[utf8]{inputenc}
\usepackage[T1]{fontenc}

\usepackage{formatting}
\usepackage{shortcuts}
\newtheorem{conj}{BGS Conjecture: Lyapunov version}
\newtheorem{conju}{BGS Conjecture: ETH version}
\newtheorem{alg}{Metropolis algorithm}

\begin{document} 
\begin{titlepage}
\begin{center}
\phantom{ }
\vspace{3cm}

{\bf \Large{Two types of quantum chaos: testing the limits of the\\[7 pt] Bohigas-Giannoni-Schmit conjecture}}
\vskip 0.5cm
Javier M. Magan${}^{\dagger, 1}$, Qingyue Wu${}^{\ddagger,2}$
\vskip 0.05in
\small{  ${}^{1}$ \textit{Instituto Balseiro, Centro At\'omico Bariloche}}
\vskip -.4cm
\small{\textit{ 8400-S.C. de Bariloche, R\'io Negro, Argentina}}

\vskip -.10cm
\small{${}^{2}$ \textit{David Rittenhouse Laboratory, University of Pennsylvania}}
\vskip -.4cm
\small{\textit{ 209 S.33rd Street, Philadelphia, PA 19104, USA}}

\begin{abstract}
There are two types of quantum chaos: eigenbasis chaos and spectral chaos. The first type controls the early-time physics, e.g. the thermal relaxation and the sensitivity of the system to initial conditions. It can be traced back to the Eigenstate Thermalization Hypothesis (ETH), a statistical hypothesis about the eigenvectors of the Hamiltonian. The second type concerns very late-time physics, e.g. the ramp of the Spectral Form Factor. It can be traced back to Random Matrix Universality (RMU), a statistical hypothesis about the eigenvalues of the Hamiltonian. The Bohigas-Giannoni-Schmit (BGS) conjecture asserts a direct relationship between the two types of chaos for quantum systems with a chaotic semiclassical limit. The BGS conjecture is challenged by the Poissonian Hamiltonian ensembles, which can be used to model any quantum system displaying RMU. In this paper, we start by analyzing further aspects of such ensembles. On general and numerical grounds, we argue that these ensembles can have chaotic semiclassical limits. We then study the Poissonian ensemble associated with the Sachdev-Ye-Kitaev (SYK) model. While the distribution of couplings peaks around the original SYK model, the Poissonian ensemble is not $k$-local. This suggests that the link between ETH and RMU requires of physical $k$-locality as an assumption. We test this hypothesis by modifying the couplings of the SYK Hamiltonian via the Metropolis algorithm, rewarding directions in the space of couplings that do not display RMU. The numerics converge to a $k$-local Hamiltonian with eigenbasis chaos but without spectral chaos. We finally comment on ways out and corollaries of our results. 
\end{abstract}
\end{center}

\small{\vspace{4 cm}\noindent
${}^{\dagger}$javier.magan@cab.cnea.gov.ar \\
${}^{\ddagger}$ aqwalnut@sas.upenn.edu
}

\end{titlepage}

\setcounter{tocdepth}{2}

{\parskip = .4\baselineskip \tableofcontents}
\newpage

\section{Introduction}\label{I}

Ever since the seminal papers \cite{10.2307/1969956,osti_4801180,PhysRevLett.52.1}, it is widely believed that sufficiently complex and chaotic physical systems have a quantized spectrum whose fine-grained properties are well approximated by Random Matrix Theory \cite{bookhaake,Guhr_1998,akemann2011oxford}. This insightful expectation of Random Matrix Universality (RMU) has been numerically verified in many systems. General arguments in favor have been described, e.g the sigma model approach \cite{Andreev:1996tw,Andreev:1996twa,Altland:2014wna} and the semiclassical theory of periodic orbits \cite{Gutzwiller,1985RSPSA.400..229B,jh}. For these and other reasons, RMU is in practice used as a definition of quantum chaos. Below we denote this facet of quantum chaos as ``spectral chaos'' or RMU. Due to the energy-time uncertainty relation, RMU shows its imprints in the very late-time physics, e.g. the ramp of the Spectral Form Factor \cite{bookhaake,Guhr_1998,akemann2011oxford}.

There is another facet of quantum chaos. This is the sensitivity of the system to initial conditions. In the classical limit, it can be characterized by the classical Lyapunov spectrum. In the quantum regime, it can be characterized by the growth of commutators  and Out-of-Time-Ordered-Correlation functions (OTOC) \cite{Larkin1969QuasiclassicalMI,Maldacena:2015waa}.\footnote{In billiard dynamics, one can approach the classical Lyapunov spectrum by the expectation value of the commutator of posititon $q(t)$ and momentum $p$ operators in a coherent state. In certain many body quantum systems, it can be more natural to use the thermal ensemble. But computing the commutator in the thermal ensemble of more generic operators might give zero due to cancellations. One can solve this problem by computing the square of the commutator, leading to OTOC. In a semiclassical limit in a coherent state, the average of the square is the square of the average, so OTOC captures as well the semiclassical physics \cite{Larkin1969QuasiclassicalMI,Maldacena:2015waa}.} This facet of chaos can be traced back to the Eigenstate Thermalization Hypothesis (ETH) \cite{PhysRevA.43.2046,PhysRevE.50.888,PhysRevE.99.042139}. This is a statistical hypothesis about the eigenvectors of a Hamiltonian. Equivalently, it is a statistical hypothesis about the unitary matrix that diagonalizes the Hamiltonian in the observable basis.\footnote{ETH is usually formulated as a statistical hypothesis for the observables in the energy basis. But we can also think it is a proposal for the statistics of eigenvectors in the observable basis. At any rate, it is transparent it has nothing to do a priori with spectrum statistics.} The mean, variance, and higher point functions in the ETH ensemble permit the computation of the one, two, and higher point functions respectively of time-dependent observables in states of interest, e.g. the thermal ensemble. This notion of chaos then controls early-time physics, e.g. thermal relaxation and sensitivity to initial conditions. Below we call this face of quantum chaos ``eigenbasis chaos''. 

The Bohigas-Giannoni-Schmit (BGS) conjecture \cite{PhysRevLett.52.1} asserts a deep relation between these two types of chaos. In the past, part of the problem with this conjecture has been the lack of a precise formulation. For this article, we start by formulating it as:

\begin{conj}\label{BGS1}
    {\sl A quantum theory whose correlation functions thermalize and OTOC shows Lyapunov growth, i.e. displaying eigenbasis chaos, also displays spectral chaos.}
\end{conj}

Instead of focusing on classical chaos in its strongest versions, one could focus in softer more general versions, such as thermalization and ergodicity. In this vein there is a natural stronger formulation of the BGS conjecture in terms of ETH

\begin{conju}\label{BGSE1}
    {\sl A quantum theory displaying ETH also displays spectral chaos.}
\end{conju}

In this article, we shed new light on these modern formulations of the BGS conjecture. The first objective will be to refine their assumptions. Once this has been clarified, the second more ambitious goal is to test its limits.

We start in section (\ref{II}) by arguing that, at least within this broad formulations, the conjecture does not hold. To this end we continue the analysis of the Poissonian Hamiltonian ensembles proposed in Ref. \cite{SpreadC,Balasubramanian:2023kwd}. In such references, to analyze the differences between systems with RMU and systems without RMU, it was noticed that for any theory with spectral-chaos (e.g. Random Matrix Theory, the Sachdev-Ye-Kitaev model (SYK) \cite{kitaev,sachdev}, local systems such as spin chains, etc), one can build an associated Hamiltonian ensemble with Poisson statistics. Equivalently, for any theory satisfying ETH and RMU, one can build an ensemble of Hamiltonians satisfying ETH but not RMU. The examples constructed satisfied ETH and Poisson statistics.\footnote{These Poissonian Hamiltonian ensembles have since then been studied further from different perspectives in \cite{Erdmenger:2023shk,Nandy:2024zcd,Gu:2024hmj,Lee:2024zvj,dssyk}.} But one can choose any spectral statistics alike. In the reverse direction, as we further comment below, one can start from an integrable Hamiltonian with neither ETH nor RMU, such as Anderson localized models, and construct an associated ensemble of Hamiltonians without ETH but with RMU. Ref. \cite{SpreadC,Balasubramanian:2023kwd} then showed that these new Poissonian ensembles accurately approximate the exact dynamics of the system until very long times. The difference between the original theory and the associated Poissonian ensemble dies in the thermodynamic limit.\footnote{Other recent examples with a dissociation between ETH and RMU are systems with large degeneracies separated by gaps, such as those appearing in supersymmetric systems at low temperatures \cite{Lin:2022zxd,Chen:2024oqv}. This has been termed ``BPS chaos''. These systems do not thermalize though, since the Hamiltonian is zero. They concern an interesting basis mismatch between the ground state space and simple operators at finite energies.}

Below we review the definition of these Poissonian ensembles and compute their spectral statistics. By construction, these statistics only depend on the average density of states of the theory. We then argue that spectral modifications of this sort cannot change early-time physics. In particular, they cannot change the early time eigenbasis chaos.\footnote{The fact they still satisfy ETH will be ovious.}We will provide a general analytical argument by bounding the difference between the correlation functions of the original theory and the Poissonian theory. We will also verify this numerically. Since the semiclassical limit and the Lyapunov spectrum are controlled by low-point correlation functions, this shows the Poissonian Hamiltonian ensembles are counterexamples to the BGS in its first formulation (\ref{BGS1}). We will end section (\ref{II}) by discussing the typicality of these ensembles in the space of Hamiltonians.

The previous discussion stems from the fact that ETH and RMU are, in principle, uncorrelated. There is no a priori relation between eigenvectors and eigenvalues. One can modify eigenvectors without affecting the eigenvalues and vice versa. So there must be a physical assumption that the first formulation (\ref{BGS1}) of the BGS conjecture is missing. The most natural idea that comes to mind is $k$-locality. More precisely, physical systems are described by Hamiltonians involving interactions of $\mathcal{O}(1)$ degrees of freedom in the thermodynamic limit. Indeed, physical Hamiltonians typically have two-, three-, or four-body interactions.\footnote{From the purely quantum mechanical perspective, this fact seems somewhat arbitrary. In local systems, especially in Quantum Field Theory (QFT), one can argue that terms involving many operators are irrelevant to the long-distance description.} In section (\ref{III}) we will show that typical draws from the Poissonian Hamiltonian ensembles are not $k$-local. This is so even if they were constructed starting from $k$-local Hamiltonians. We provide a general argument and show it for SYK numerically. In the SYK scenario, while the Poisson-ize version peaks around the original SYK, it has non-perturbative fermion string tails of arbitrary size. These results suggest the following slight but insightful modification to the formulation of the BGS conjecture

\begin{conj}\label{BGS2}
    {\sl A quantum theory driven by a $k$-local Hamiltonian whose correlation functions thermalize and OTOC shows Lyapunov growth, i.e. displaying basis-chaos, also displays spectral chaos.}
\end{conj}

As before, we can strengthen the conjecture as

\begin{conju}\label{BGSE2}
    {\sl A quantum theory driven by a $k$-local Hamiltonian displaying ETH also displays spectral chaos.}
\end{conju}

In section (\ref{IV}) we explore the limits of these formulations. Now we cannot simply modify the spectrum of the theory since there is an intricate relation between eigenvalues and $k$-locality. This is precisely controlled by ETH. Indeed, most changes in the spectrum will lead to Hamiltonians which are not $k$-local. The only way to make sure we remain $k$-local is to explore the coupling space of $k$-local terms. However, we do not know what directions we should choose to erase RMU. A possibility then is to blindly explore the coupling space, rewarding directions in which we destroy RMU. This can be approached numerically via the Metropolis algorithm. Following the BGS conjecture (\ref{BGS2}), this Markovian process should end in a distribution with no spectral-chaos, but with no basis-chaos either (e.g. some glassy or integrable phase). While we were expecting that result, the end-point still shows early time chaos and thermalization. Although not in the thermodynamic limit, this suggests that the second formulation might not hold either.

We end in section (\ref{V}) with an open discussion of different applications, problems, and open directions. This includes applications in quantum gravity and a further formulation of the conjecture that still survives the present analysis.

\section{Poissonian Hamiltonian ensembles}\label{II}

We continue the analysis of the Poissonian Hamiltonian ensembles proposed in Ref. \cite{SpreadC,Balasubramanian:2023kwd}. Such works noticed that
in any theory with a dense spectrum, it is natural to define an associated ensemble of Hamiltonians with a spectrum displaying the same average density of states, but with Poissonian statistics.\footnote{Note we can modify the universality class at will. This is done on \cite{dssyk} for the case of SYK. But for the problem considered in this article, it is more interesting to consider the Poissonian scenario.} This ensemble accurately approximates the original Hamiltonian until very long times. This is possible even if the original theory displays spectral chaos, both for random Hamiltonian ensembles and for particular Hamiltonians such as lattice models.

To be precise, consider a Hamiltonian $H$ with eigenvalues $E_n$ and eigenvectors $\vert n\rangle$. If this Hamiltonian is drawn itself from an ensemble of Hamiltonians, then the erratic density of states 
\be \label{denm}
\rho(E)\equiv\frac{1}{N}\sum \delta(E-E_i)\;,
\ee
will average to a continuous function $\overline{\rho(E)}$ in the thermodynamic limit, e.g. $\overline{\rho(E)}=\frac{1}{2\pi}\sqrt{4-E^2}$ for Gaussian random matrices \cite{akemann2011oxford}. Above we have defined $N$ to be the dimension of the Hilbert space. If instead we have a particular Hamiltonian, e.g. a local lattice Hamiltonian, its spectra will fluctuate around some coarse-grained local mean.  One can find this mean by locally smoothing the spectrum, see \cite{Balasubramanian:2023kwd} for specific details and applications.

In both scenarios, we end up with a continuous function $\overline{\rho(E)}$. This is the density of states of the theory in the thermodynamic limit. Even if the original spectrum displays fine-grained properties, such as RMU, we now erase such features. More precisely we define an ensemble of Hamiltonians with the same eigenvectors but with eigenvalues drawn at random from $\overline{\rho(E)}$. This function is seen now as a probability distribution, and the new eigenvalues are located in the same ascending order. Intuitively, a typical instance of this ensemble is a Hamiltonian with the same thermodynamic density of states and eigenvectors of the original theory, but with no spectrum-correlations.\footnote{As mentioned in the introduction, we can also modify the spectrum to a different chaotic universality class. We can even take an initially integrable system, with neither ETH nor RMU and modify the spectrum to display RMU.} Mathematically, for any theory with a continuous density of states $\overline{\rho(E)}$ in the thermodynamic (or semiclassical) limit, Ref. \cite{SpreadC,Balasubramanian:2023kwd} defines an ensemble of Hamiltonians whose joint probability distribution for the eigenvalues is simply
\be\label{pois}
\rho(E_1,\ldots,E_n)=\prod_i \overline{\rho (E_i)}\;.
\ee
This defines a Poissonian Hamiltonian ensemble associated with the original theory. It erases RMT correlations if the original theory displayed them. More generally we can define different ensembles with different joint probability distributions for the eigenvalues. The only constraint amounts to fixing the average density of states $\overline{\rho(E)}$. Since only modifies the spectrum, if the system originally satisfied ETH, the Poissonian ensemble will still satisfy ETH. They then violate the strongest formulations of the conjecture.

\subsection{Poissonian spectral statistics}

For completeness and later use, we now compute the spectral statistics of Poissonian ensembles. The one-point function is given by $\overline{\rho (E)}$, as it should. The two-point function can be decomposed into connected and disconnected components giving
\be\label{twop}
\overline{\rho(E)\rho(E')}=\left(1-\frac{1}{N}\right)\overline{\rho(E)}\,\overline{\rho(E')}+\frac{1}{N}\overline{\rho(E)}\delta(E-E')\;.
\ee
From this correlation function we compute the Spectral Form Factor \cite{bookhaake,Guhr_1998,akemann2011oxford}. This is defined as
\be \label{SFF}
\textrm{SFF}_{\beta}(t)\equiv Z_{\beta-it}Z_{\beta+it}=\sum_{mn}e^{-(\beta-it)E_m}e^{-(\beta+it)E_n}\;.
\ee
On average and in the thermodynamic limit, the SFF is the Fourier transform of the two-point correlation function~(\ref{twop}). We obtain
\be\label{asff}
\overline{\text{SFF}_\beta(t)}=\vert\overline{Z_{\beta-it}}\vert^2+Z_{2\beta} \;.
\ee
This first term corresponds to the disconnected contribution. It provides the universal decay of the SFF at early times. It is controlled by the average density of states. For example, for Gaussian random matrices, the decay of this term is controlled by a Bessel function. This decay cannot last forever for systems with a finite number of degrees of freedom. For systems with an erratic spectrum and no degeneracies, this follows from the time average of the exact SFF, eq (\ref{SFF}), leading to the value $Z_{2\beta}$. Although Poissonian spectra might display degeneracies, the system is still finite. It should plateau as well. The second term in (\ref{asff}) says that the same plateau applies to the Poissonian ensemble. Therefore, Poissonian degeneracies are not strong enough to disturb the plateau. The saturation of the SFF was studied in \cite{Cotler:2016fpe} as a simplified version of Maldacena's avatar of the information paradox in AdS/CFT \cite{Maldacena:2001kr}. As explained there, obtaining the plateau in RMT requires control over non-perturbative corrections (in an expansion in terms of the Hilbert space dimension). For the Poissonian Hamiltonian ensembles, the plateau is obtained readily as the first and only correction. We will further comment on quantum gravity applications in the discussion section.

Similarly, we can also compute higher moments of the Poissonian eigenvalue distribution. Let's start with the three-point function. This is defined as
\be 
\overline{\rho(E_1)\rho(E_2)\rho(E_3)}=\frac{1}{N^3}\int\prod_{l=1}^{N} d\lambda_l \rho(\lambda)\sum_{ijk}\delta (E_1-\lambda_i) \delta (E_2-\lambda_j) \delta (E_3-\lambda_k)\;.
\ee
For $i\neq j\neq k$ we have $\frac{N(N-1)(N-2)}{N^3}$ terms, each contributing $\overline{\rho(E_1)}\,\overline{\rho(E_2)}\,\overline{\rho(E_3)}$. These are fully disconnected contributions. For $i\neq j= k$ we have $\frac{N(N-1)}{N^3}$ terms, each contributing $\frac{1}{N}\,\overline{\rho(E_1)}\,\overline{\rho(E_2)}\delta (E_2-E_3)$. Similar contributions go for $i= j\neq k$ and $i= k\neq j$. These contributions connect two densities and leave one factorized. Finally for $i=j=k$ we have $N$ terms, each contributing $\frac{1}{N^2}\,\overline{\rho(E_1)}\,\delta (E_1-E_2)\delta (E_1-E_2)$. This is the fully connected contribution.

The structure of the three-point function paves the way towards the generic n-point correlation function. We need to separate the sum into different types of terms. These are classified by which indices are repeated. For each type, we need to count the number of terms and then evaluate its contribution. More concretely, suppose we need to compute
\be \label{npoint}
\overline{\rho(E_1)\cdots\rho(E_n)}=\frac{1}{N^n}\int\prod_{l=1}^{N} d\lambda_l \rho(\lambda)\sum_{i_1\cdots i_m}\delta (E_1-\lambda_{i_1})\cdots \delta (E_m-\lambda_{i_m})\;.
\ee
A generic term in the sum has 
\be
i_1=i_2=\cdots=i_{l_1}\neq i_{l_{1+1}}=i_{l_{1+2}}=\cdots =i_{l_{1}+l_2}\neq \cdots \neq i_{l_{1}+l_2+\cdots+l_r+1}=\cdots = i_{l_{1}+l_2+\cdots+l_r+l_{r+1}}\;,
\ee
with $\sum_{i=1}^{r+1}l_i=m$. There are
\be 
\binom{N}{l_1}\binom{N-l_1}{l_2}\binom{N-l_1-l_2}{l_3}\cdots \binom{N-l_1-\cdots-l_{r-1}}{l_r}
=\frac{N!}{\prod_i l_i!}\;,
\ee
terms of this form, each contributing
\be 
\overline{\rho(E_1)}\left(\prod_{i=2}^{l_1}\delta(E_1-E_i)\right)\,\overline{\rho(E_{l_{1+1}})}\left(\prod_{j=l_1+2}^{l_2}\delta(E_{l+1}-E_j)\right)\cdots \overline{\rho(E_{\sum_{s=1}^{r}l_s})}\left(\prod_{p=\sum_{s=1}^{r}l_s+2}^{\sum_{s=1}^{r+1}l_s}\delta(E_1-E_p)\right)\;.\nonumber
\ee
Adding these contributions for all types of terms we compute (\ref{npoint}).

\subsection{Lyapunov exponents and ETH with Poisson statistics}

What are the dynamical consequences of this drastic change in the Hamiltonian spectrum? Ref. \cite{SpreadC,Balasubramanian:2023kwd} showed that several time-dependent quantities were well approximated by the associated Poissonian ensembles. An obvious example is the SFF itself. This is controlled by the disconnected piece of the density-density correlation function for very long times. We now argue that this is also the case for low-point correlation functions of simple observables. We provide a general argument and then numerically verify this expectation for SYK. Since these correlation functions control the semiclassical limit, this will show that the Poissonian ensembles violate the BGS conjecture in its first formulation (\ref{BGS1}).\footnote{As we mentioned above, the fact they violate the BGS conjecture in its ETH first formulation is trivial.}

For a given theory with Hamiltonian $H$, consider a general thermal correlation function of the form
\be  
    \textrm{Tr} \left( e^{-\beta_1H}\,\mathcal{O}_1(t_1)\,e^{-\beta_2H}\mathcal{O}_2(t_2)\,\cdots\,e^{-\beta_nH}\mathcal{O}_n(t_n)\right)=\textrm{Tr} \left( \prod_i e^{-\beta_iH}e^{iHt_i}\mathcal{O}_i e^{-iHt_i}\right)\;.
\ee
At constant time, if the density of states remains constant, this scales with the dimension of the Hilbert space $N$ thanks to the trace. The objective is to bound the difference between this correlator and the associated Poisson-ize version.

Let $C(E)=\int_{-\infty}^E \rho(E')dE'$ be the cumulative density of states. Then, in both the original and Poisson-ize version, the $i$-th smallest eigenvalue should be $C^{-1}(i/N)+O(1/N)$. Since the order of the eigenvalues is preserved in the new ensemble, this implies that the energy eigenvalues change by $O(1/N)$ when we Poisson-ize the Hamiltonian. Equivalently, the matrix $\delta H = H' - H$ has $O(1/N)$ eigenvalues and commutes with both $H'$ and $H$.

To leading order, the correction to the correlation function is given by a sum of terms of the form
\be
\textrm{Tr}\,\Delta\equiv\textrm{Tr} \left( \prod_{i<a} e^{-\beta_iH}e^{iHt_i}\mathcal{O}_i e^{-iHt_i}(e^{C\delta H}-1) \prod_{i\geq a} e^{-\beta_iH}e^{iHt_i}\mathcal{O}_i e^{-iHt_i}\right)\:.
\ee
Consider first the case where $H-E$ is positive semidefinite, where $E$ is some lower bound for the energy, and the $O$'s have eigenvalues bounded in some fixed range $[-A,A]$ as $N$ increases. In this case, the operator $\Delta$ within the trace has eigenvalues $\lambda_{\Delta}$ bounded by
\be
    \lambda_{\Delta} < e^{-\sum_i \beta_i E}A^n|e^{C\delta H}-1|\;,
\ee
which decreases to zero as $N$ increases. The assumption of lower bounded $H$ and bounded $O$ is a physical one. Whenever we look at a system only through a finite energy window, we are effectively considering bounded $H,O$.

For systems with unbounded eigenvalues, it is helpful to bound this difference via a softer method. This uses the Schatten operator norm, $|A|_p=(\sum s_i(A)^p)^{1/p}$, where $s_i(A)$ are the singular values of $A$. Holder inequality for the Schatten norm gives the following bound
\bea
    \tr(\Delta)<|\Delta|_{1}&<& \left\lvert (e^{C\delta H}-1)\prod_{i\geq a} e^{-\beta_iH}e^{iHt_i}\mathcal{O}_i e^{-iHt_i}\prod_{i < a} e^{-\beta_iH}e^{iHt_i}\mathcal{O}_i e^{-iHt_i} \right\rvert_1 \nonumber\\ &<&  |e^{C\delta H}-1|_{\alpha} \left\lvert \prod_{i\geq a} e^{-\beta_iH}e^{iHt_i}\mathcal{O}_i e^{-iHt_i}\prod_{i < a} e^{-\beta_iH}e^{iHt_i}\mathcal{O}_i e^{-iHt_i} \right\rvert_\gamma
\eea
when $1/\alpha+1/\gamma=1$, with $\alpha, \gamma>0$. The first term is $O(N^{1/\alpha-1})$, while the second term scales with $O(N^{1/\gamma})$, so this norm is again suppressed to order $O(1)$.

We can provide further physical intuition by assuming ETH in the original Hamiltonian. In this case, the Poisson version also satisfies the same ETH. We then rewrite the correlation function in the energy basis
\be  
    \textrm{Tr} \left( e^{-\beta_1H}\,\mathcal{O}_1(t_1)\,e^{-\beta_2H}\mathcal{O}_2(t_2)\,\cdots\,e^{-\beta_nH}\mathcal{O}_n(t_n)\right)=\sum\limits_{j_1\cdots j_n} \prod_i e^{-\beta_i E_{j_i}} \prod_i e^{iE_{j_i}(t_i-t_{i-1})}\mathcal{O}^1_{j_1 j_2}\,\cdots\,\mathcal{O}^n_{j_n j_1}\;,
\ee
where $O^i_{ab}=\bra{E_a}O^i\ket{E_b}$ are the entries of the operator $O^i$ in the energy basis and we have made the identification $t_0=t_n$. The change in the spectrum of the Hamiltonian does not affect the statistics of the operators $\mathcal{O}^1_{j_1 j_2}\,\cdots\,\mathcal{O}^n_{j_n j_1}$. In the continuum limit, the correlation function is the integral $\int (\prod_i \rho(E_i)dE_i) f(E_i)g(E_i)$ of a smooth function $f(E_i)=\prod_i e^{-\beta_i E_{i}} \prod_i e^{iE_{i}(t_i-t_{i-1})}$, noisy functions $g(E_{j_i})=\mathcal{O}^1_{j_1 j_2}\,\cdots\,\mathcal{O}^n_{j_n j_1}$, together with the product of the average density of states $\prod_i \rho(E_i) $. The leading term in this expression can be obtained by taking the average over the noisy $g(E_{j_i})$ factors in the ETH ensemble and the product of average densities of states $\prod_i \overline{\rho(E_i)}$. This leading approximation shows no difference between RMU and Poisson distributed eigenvalues, at least until very long times.  This argument stems from the fact that corrections to the leading answer from eigenvalues statistics are exponentially suppressed in the entropy. On the other hand, higher point functions in the ETH ensemble provide larger contributions, see \cite{PhysRevE.99.042139,Jafferis:2022wez}. Therefore, while Gaussian and non-Gaussian ETH correlations need to be taken into account for the early time physics, in particular the early time chaos, spectral correlations remain subleading until very long times.

These general arguments can be verified numerically in the SYK model \cite{kitaev,sachdev}. The SYK model is an ensemble of Hamiltonians over $N$ Majorana fermions $\psi_i$, $i=0,\cdots,N-1$, with 
$\left\lbrace \psi_i,\psi_j  \right\rbrace = \psi_i\psi_j+\psi_j\psi_i=2\delta_{ij}$. It is defined as
\be
\label{hamilsyk}
H=i^{p/2}\sum\limits_{0\leq i_1<i_2<i_3 <i_4<N}\,J_{i_1i_2i_3i_4}\,\psi_{i_{1}}\psi_{i_{2}}\psi_{i_{3}} \psi_{i_{4}}\;,
\ee
where the couplings $J_{i_1i_2i_3i_4}$ are real and independently distributed Gaussian random variables, namely
\be 
\langle J_{i_1 i_2 i_3i_4} \rangle =0\,\,\,\,\,\,\,\,\,\,\,\,\,
\langle J^2_{i_1 i_2 i_3i_4} \rangle=\frac{1}{\lambda}{\begin{pmatrix}
N\\
4
\end{pmatrix} }^{-1} \mathcal{J}^2\;,
\ee
where we use $\lambda = \frac{2p^2}{N}$, and $p=4$. For this article, we restrict ourselves to $N$ such that $N \equiv 2~~(\text{mod}~4)$. This way we find the RMT universality class to be the GUE class \cite{You:2016ldz}.

To represent the Majorana fermions we use the standard Jordan-Wigner representation over $N/2$ spins. On these spins, the fermions can be represented via
\bea
\psi_{2i}&=&\sigma_z \otimes \ldots \otimes \sigma_z \otimes \sigma_x \otimes 1 \otimes \ldots \otimes 1\;,\nonumber\\
\psi_{2i+1}&=&\sigma_z \otimes \ldots \otimes \sigma_z \otimes \sigma_y \otimes 1 \otimes \ldots \otimes 1\:,\label{jw_rep}
\eea
where there are $i$ copies of $\sigma_z$, and the tensor product $\otimes$ is internally represented as a Kronecker product. There is a parity symmetry separating the system into even and odd sectors of the Hilbert space. If we consider the basis of tensor products of spin up/down states, these sectors are differentiated by the states having an even/odd number of spin-ups in the tensor product.

\begin{figure}
    \includegraphics[width=0.98\linewidth]{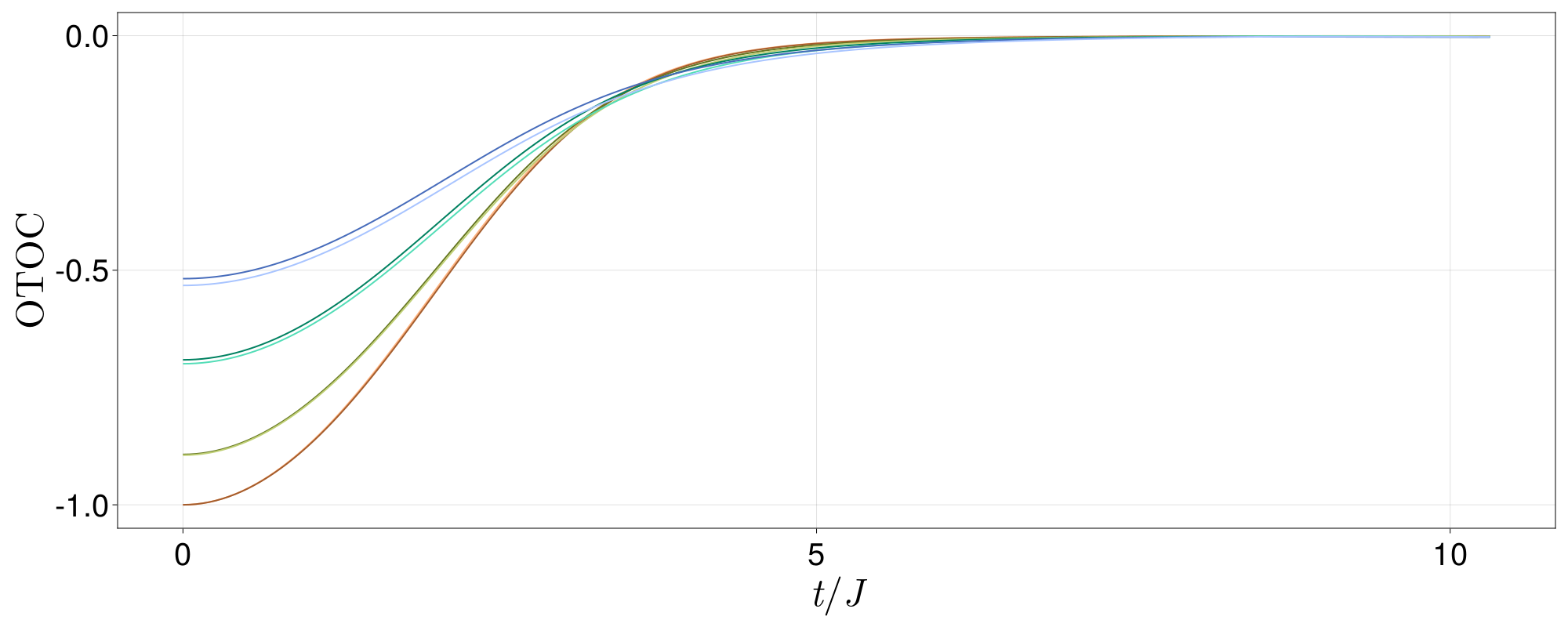}
    \caption{\label{fig-otoc} The OTOC (\ref{OTOCp}) at temperatures $\beta=0,1,2,3$ (corresponding to orange, yellow, green, and blue respectively) for a single sample of an $N=22$ SYK Hamiltonian (darker colors) and its associated Poisson-ized version (lighter colors).
    }
\end{figure}

We also consider the Poisson-ized version of the SYK Hamiltonian ensemble. Let us describe the procedure to construct it. We first create a pool of eigenvalues by taking many Hamiltonian samples from the SYK ensemble. We numerically diagonalize them and collect their eigenvalues. We then diagonalize a further copy of SYK, writing the Hamiltonian as $H=U^\dagger D U$, and split it into its even and odd sectors. In each sector, we replace the eigenvalues with random samples from the pool to create Poisson-distributed eigenvalues $D'$. The new eigenvalues are sorted in the natural order. In each sector, the lowest eigenvalue of $D$ is at the same entry as the lowest eigenvalue of $D'$, the second lowest eigenvalue in $D$ is at the same entry as the second-lowest eigenvalue of $D'$, and so on. Our final Poisson-ized Hamiltonian is computed via $H'=U^\dagger D' U$.

In Fig. \ref{fig-otoc} we plot the out of time order correlation function\footnote{We follow the standard conventions of Ref. \cite{Maldacena:2015waa}.} 
\be\label{OTOCp}
    \text{OTOC}_{\beta}(t) = \frac{1}{Z_\beta}\text{Tr}\left(e^{-\frac{\beta}{4} H}\psi_1(t) e^{-\frac{\beta}{4} H}\psi_2(0)e^{-\frac{\beta}{4} H}\psi_1(t) e^{-\frac{\beta}{4} H}\psi_2(0)\right)\;,
\ee
for an instance of \eqref{hamilsyk}, together with the OTOC for the modified Poisonnian Hamiltonian ensemble. In the latter, we have used a pool of $256$ SYK Hamiltonians. We see that both correlation functions match well for times much smaller than the dimension of the Hamiltonian. 

Summarizing, we conclude that thermalization, ETH, and Lyapunov growth, i.e. the early time or eigenbasis chaos, are in principle dissociated from RMU, i.e. from spectral chaos. Interestingly, this also works in the reverse direction. We can start from an integrable model without Lyapunov growth, ETH, or RMU, and then modify the spectrum to display spectral chaos. After this modification, the localization properties of eigenstates are unchanged.

\subsection{Typicality of Poissonian Hamiltonian ensembles}

We have argued in several manners that the Poissonian Hamiltonian ensembles of Ref. \cite{SpreadC,Balasubramanian:2023kwd} demonstrate that the first formulations of the BGS conjecture are not valid. An immediate reaction is that these Hamiltonians might not be typical in the space of Hamiltonians. Indeed, typicality arguments were used in the past as evidence of the BGS conjecture, see \cite{Guhr_1998}. This expectation turns out to be incorrect. We now expand on this.

The notion of (non)-typicality strongly depends on the physical features we fix at the start. The Jaynes principle of maximal ignorance \cite{PhysRev.106.620} provides the right guide for this discussion. This principle instructs us to maximize entropy given a set of expectations. In this vein consider a Hamiltonian in an $N$-dimensional Hilbert space. This is a $N\times N$ matrix. In certain natural (or computational) basis, we could decide the entries $H_{ij}$ have prescribed means and variances. Applying the principle of maximum ignorance we arrive at one of the Gaussian ensembles of RMT. Hamiltonians drawn from this ensemble typically display RMU. We conclude that, if the Hamiltonian entries on a certain basis are Gaussian random numbers, the Poissonian Hamiltonian ensembles discussed above are highly atypical.

But in certain scenarios, there might be no preferred Hamiltonian basis in which the entries are Gaussian random numbers. Indeed, in some scenarios, it could be more natural (or less biased) to fix the average density of states $\overline{\rho (E)}$. Jaynes's principle of maximum ignorance then leads to the Poissonian Hamiltonian ensemble, defined by eq. (\ref{pois}). Of course, Hamiltonians drawn from this ensemble typically have a spectrum displaying Poissonian statistics and not RMU. Hamiltonians displaying RMU are highly atypical in this ensemble.

Interesting and intriguing cases where fixing the density of states (instead of matrix entries on a particular basis) seems most natural are theories of quantum gravity and CFTs. We will further expand on these issues in the discussion section. As for now, without further assumptions, we remark that by fixing aspects of the density of states, the only way to get RMU is by fixing by hand the two-point function to be the universal RMT answer itself.

\section{Poissonian Hamiltonians ensembles are not $k$-local}\label{III}

We have shown there are general families of Hamiltonians invalidating the first formulation of the BGS conjecture (\ref{BGS1}). Given the extensive zoo of realistic models where this conjecture has been verified, see \cite{bookhaake,Guhr_1998,akemann2011oxford} for reviews, this suggests this formulation lacks some further physical assumption.

\begin{figure}
    \includegraphics[width=0.98\linewidth]{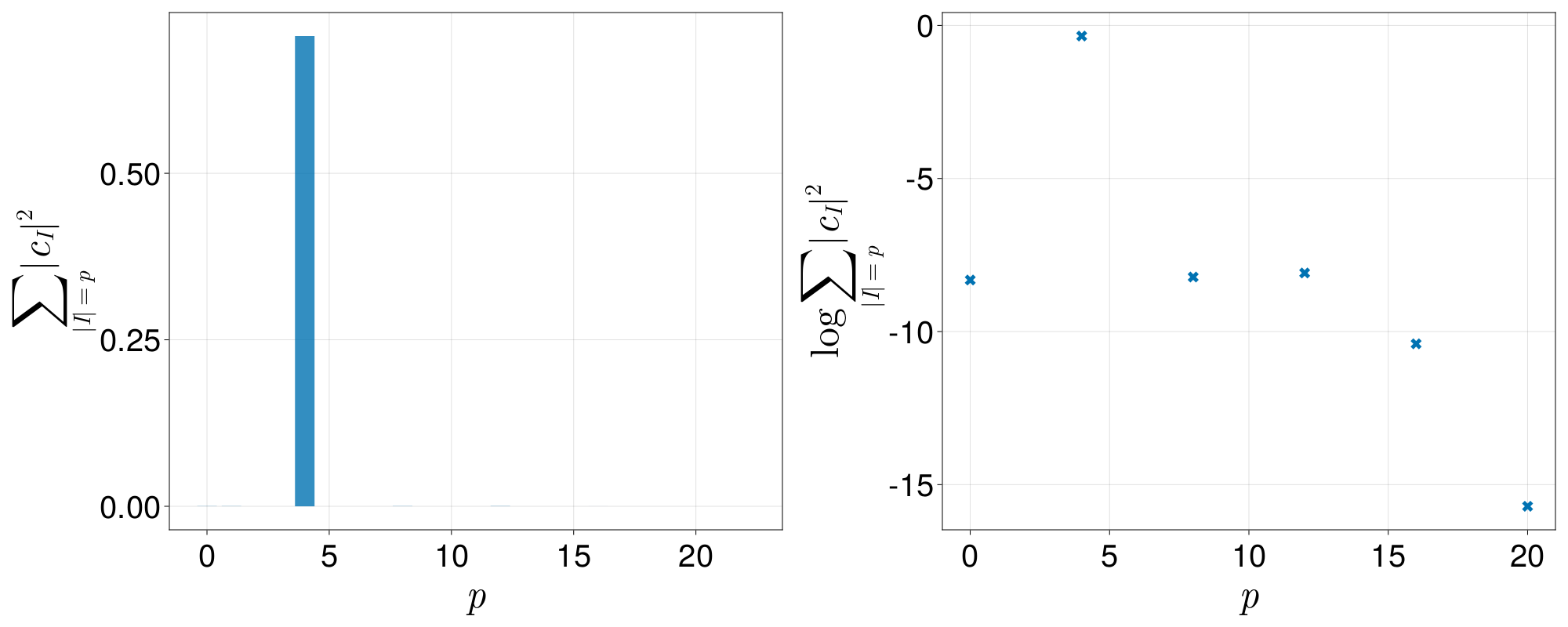}
    \caption{\label{fig-opsizes} The squared magnitudes of the $|I|$-fermion parts of the ``Poissonized'' Hamiltonian, averaged over $64$ instances of SYK hamiltonians.
    }
\end{figure}

The first idea that comes to mind is $k$-locality. In a spin system, an operator $\mathcal{O}$ is said to be $k$-local if it can be expressed as a sum of terms where each term contains products of no more than $k$ different spin operators. In the Majorana fermion scenario, $\mathcal{O}$ is $k$-local if it can be expressed as
\be 
\mathcal{O}=\sum\limits_{p=1}^{k}\, \sum_{i_1<\cdots <i_p} c_{i_1 \cdots i_p} \psi_{i_1}\ldots \psi_{i_p}\;.
\ee
Realistic systems have Hamiltonians that are at most $4$-local. This suggests to add to the first formulation (\ref{BGS1}) the assumption that the Hamiltonian is $k$-local, with $k\sim\mathcal{O}(1)$ in the thermodynamic or semiclassical limit,\footnote{We will comment on the QFT case in the discussion section.} leading to the second formulation of the BGS conjecture (\ref{BGS2}). In an abuse of language, below we say a Hamiltonian is $k$-local if it satisfies this assumption. The question we need to address concerns the $k$-locality of the Poissonian Hamiltonian ensembles.

We now argue that these ensembles are not $k$-local. Let us call $H$ to the original $k$-local Hamiltonian, and $H'$ to its Poisson-ize version. In the numerics below $H$ will be that of SYK. Both Hamiltonians are diagonalized by the same unitary matrix $U$, namely $H=U^\dagger DU$ and $H'=U^\dagger D' U$, with $D$ and $D'$ the diagonal matrix of eigenvalues of each Hamiltonian. The difference between the two is $\delta H = U^\dagger (D'-D) U$. Due to the sorted eigenvalues with the same density of states, the entries in $(D'-D)$ are generally of order $2^{-N/2}$ (in the Majorana fermion scenario). Calling $v_i$ to the eigenvectors of $H,H',\delta H$ and $d_i$ to the eigenvalues of $\delta H$ we have $\delta H=\sum_i d_i v_i v_i^{\dagger}$. So after conjugation by the unitary $U$, each entry in $(D'-D)$ contributes with $d_i v_i v_i^\dagger$ to $\delta H$. The $v_i$'s are all orthogonal, so $\tr(v_i v_i^\dagger v_j v_j^\dagger)=0$ (for $i\neq j$). This means the $2^{N/2}$ different contributions $d_i v_i v_i^{\dagger}$ are all in orthogonal directions. Therefore $\delta H$ has a Frobenius norm\footnote{We remind the Frobenious norm is defined as $\Vert A \Vert_F\equiv \sqrt{\textrm{Tr}(AA^T)}$.} of order $2^{-N/4}$. This in turn implies the new Hamiltonian is not $k$-local and it must contain non-perturbative, exponentially suppressed in the dimension of the Hilbert space, fermion tails of all sizes.

We can check this numerically by decomposing the new Hamiltonian $H'$ into its $p$-fermion components
\be
H' = \sum\limits_p\, \sum_{i_1<\cdots <i_p} c_{i_1 \cdots i_p} \psi_{i_1}\ldots \psi_{i_p}\equiv \sum\limits_p\, \sum_{I=p} c_{I} \psi_{i_1}\ldots \psi_{i_p}\;,
\ee
where $I$ is just a shorthand notation for the set of indices at some size $p$. This decomposition can be done efficiently by first decomposing the Hamiltonian into tensor products of Pauli matrices. This takes advantage of the fact that
\be
2\begin{bmatrix}H_1&H_2\\H_3&H_4\end{bmatrix}=1\otimes(H_1+H_4)+\sigma_z\otimes(H_1-H_4)+\sigma_x\otimes(H_2+H_3)+i\sigma_y\otimes(H_2-H_3)\;.
\ee
We can iterate this process on the four smaller Hamiltonian blocks until we arrive at an expression in terms of tensor products of spins. Using \eqref{jw_rep}, we can map this to the coefficients of the Majorana operators via a permutation. Multiplication by an odd number of $\sigma_z$'s at any position swaps $1$ and $\sigma_z$, and $\sigma_x$ and $\sigma_y$. An extra $\sigma_z$ implies both $\psi_{2i}$ and $\psi_{2i+1}$ were present in the product. We can work from right to left in this manner. This gives an algorithm that finds the coefficients $c_{i_1 \cdots i_p}$ in $O(n^2\log n)$ for a $n\times n$ matrix (for example $n=2^{N/2}$ in our case).

\begin{figure}
    \includegraphics[width=0.98\linewidth]{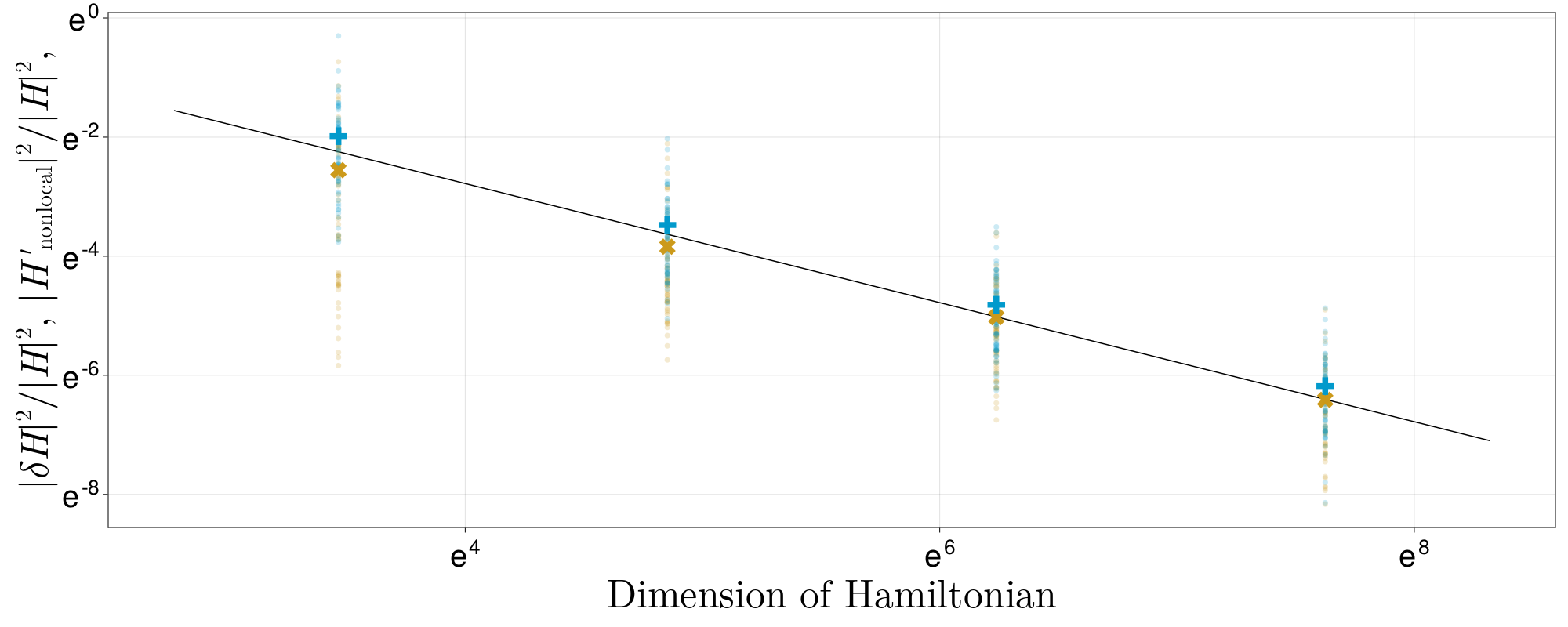}
    \caption{\label{fig-nonlocalpart} The norm of the non-local part of the ``Poisson-ized'' $N=10,14,18,22$ SYK Hamiltonian (blue dots are individual samples, their average at the blue x), and the difference from the original SYK Hamiltonian (orange dots are individual samples, their average at the orange $+$), all divided by the norm of the whole ``Poisson-ized'' Hamiltonian. The black line has slope $-1$ for reference.
    }
\end{figure}

Using this algorithm, in Fig. \ref{fig-opsizes} we compute the magnitude of the $p$-fermion components of the Poisson-ized Hamiltonian $H'$. This plot shows that $H'$ now contains terms of arbitrary size.\footnote{The different value for each size corresponds to the different number of fermions at each size, and not to a hidden size-dependent structure. The only exception is the term corresponding to the identity.} Further, in Fig. \ref{fig-nonlocalpart} we plot the magnitude of the ``non-local part''
\be 
 H'_\text{non-local} \equiv \sum_{p>4} \sum_{|I|=p} c_I \psi_{i_p}\psi_{i_p}\ldots \psi_{i_p}
\ee
of the Hamiltonian. From the plot, we conclude that the size of this correction is of order $2^{-N/4}$, as we argued on general grounds above.

\begin{figure}
    \includegraphics[width=0.98\linewidth]{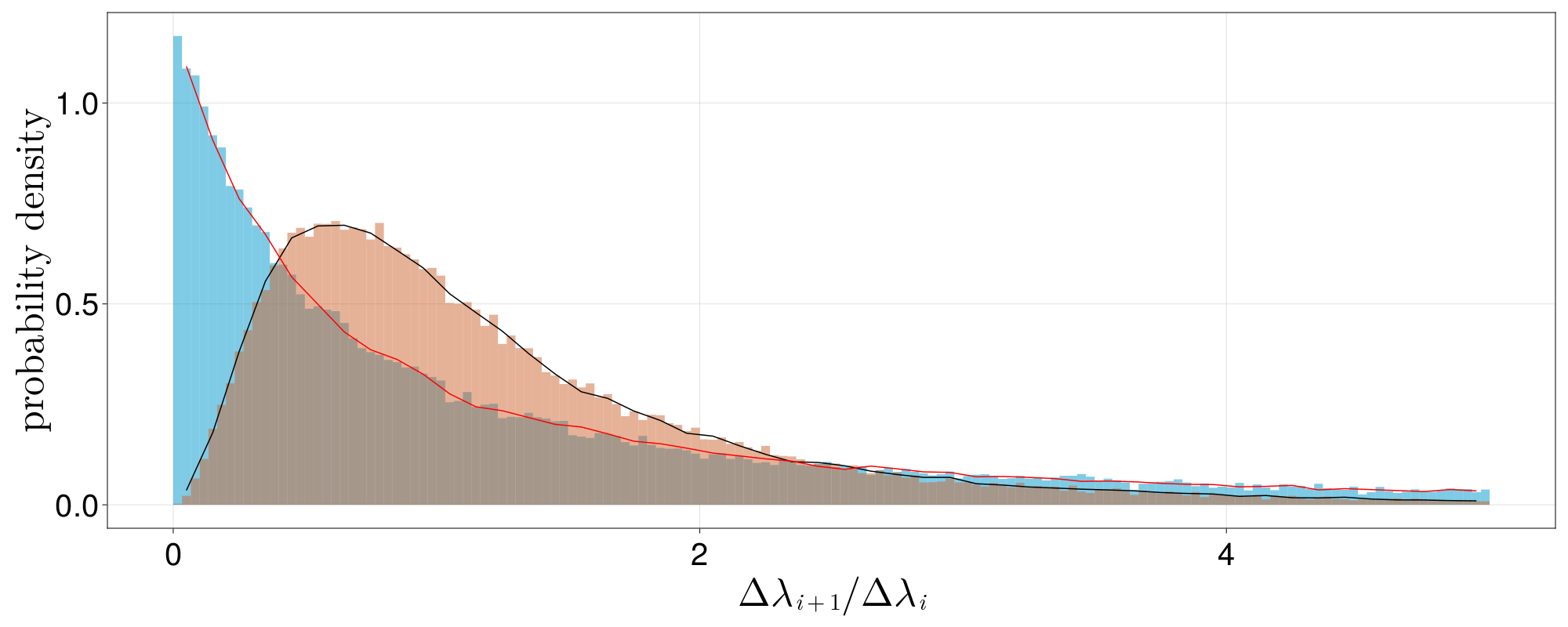}
    \caption{\label{fig-rmtcheck} A histogram (probability density norm) of successive eigenvalue gaps of $64$ samples each of: Poisson/independently sampled eigenvalues (red line), normal $N=22$ SYK (black line), ``Poisson-ized'' $N=22$ SYK (blue histogram), and ``Re-localized'' $N=22$ SYK Hamiltonians.
    }
\end{figure}

Albeit this correction might seem small, it happens to play a very significant role. For example, let us ``re-localize'' the Hamiltonian by removing the non-local part. This defines a new Hamiltonian
\be
H_\text{local}' = \sum_{p\leq 4} \sum_{|I|=p} c_I \psi_{i_p}\psi_{i_p}\ldots \psi_{i_p},
\ee
The Poissonian Hamiltonian $H'$ had by construction a Poissonian distribution for the eigenvalue gap ratios, i.e for the distribution of $\frac{\lambda_{i+1}-\lambda_i}{\lambda_i-\lambda_{i-1}}$ where $\lambda_{i-1},\lambda_{i},\lambda_{i+1}$ are nearest successive eigenvalues of the Hamiltonian. In contrast, as we show in Fig. \ref{fig-rmtcheck}, $H_\text{local}'$ no longer has a Poisson distribution and becomes distributed like a random matrix (the Wigner surmise) once again. 

This could be anticipated for the following reason. Especially at large $N$, the change $\delta H'$ will be mostly in non $k$-local directions. Even ignoring the directions of the eigenvectors, the $2^{N/2}$ directions represented in $\delta H'$ will cover exponentially more (in operator space) than the $O(N^4)$ $k$-local degrees of freedom covered by $k$-local Hamiltonians. Very little change is made to the local part of the Hamiltonian in the process of Poissonization.

\section{Poisson Distributed Eigenvalues with Exact $k$-Locality}\label{IV}

Previous considerations have led us to the second formulation of the BGS conjecture (\ref{BGS2}). This includes the assumption of $k$-locality for the Hamiltonian. We now seek to explore the limits of this formulation. Notice we cannot simply perturb the spectrum of the theory. As we showed above, most changes lead to Hamiltonians which are not $k$-local. There is an intricate relation between eigenvalues and $k$-locality which is ultimately controlled by ETH in quantum theories satisfying it.

Then, the only route we can think of concerns the exploration of the coupling space of $k$-local interactions. This can be approached numerically via sampling, i.e. via a Metropolis-type algorithm. A Hamiltonian drawn from standard Random Matrix Theory with potential $V(E)$ has eigenvalues distributed according to 
\be
    \rho(\lambda)=e^{-N\sum V(\lambda_i)} \prod_{i<j} \lvert \lambda_i-\lambda_j\rvert^{\beta_D}\;.
\ee
The eigenvalues of the $4$-local SYK Hamiltonians have a joint distribution that is well approximated by this RMT joint-eigenvalue-distribution, once we choose a suitable potential function $V(x)$. More precisely, this potential is chosen so that the leading density of states of the RMT is equal to the leading density of states of SYK. Equivalently, the nearest neighbor distribution in SYK is well approximated by the Wigner surmise for the appropriate universality class. The probability that two consecutive eigenvalues are separated by $\Delta E$ goes to zero as $\Delta E$ goes to zero.

These observations suggest the following idea. Define $\lambda(J)$ to be the eigenvalues of a SYK Hamiltonian with couplings $J$.  If instead of sampling the SYK couplings $J$ from a Gaussian distribution $e^{-kJ^2}$, we were to sample them from a distribution proportional to
\be 
P(J)\propto \frac{e^{-kJ^2}}{\rho(\lambda(J))}\;,
\ee
we would in principle be able to create $k$-local Hamiltonians that have a flat distribution of eigenvalues. Intuitively, this potential probability distribution for the couplings gives high probability to situations that were given low probability in the original Gaussian distribution. In particular, as already mentioned, the Gaussian probability distribution gives a small probability to sufficiently small energy differences, while the new distribution gives a high probability to those. One can see this more precisely if instead of the SYK coupling we use this for the entries of a Gaussian Random matrix. Then dividing by the joint probability distribution would provide the mentioned flat distribution.

There are some theoretical issues with this idea. The first problem is that $k$-local Hamiltonians cannot have exactly a RMT joint-eigenvalue distribution. The second is that $\frac{1}{\rho(\lambda(J))}$ may not integrate into a distribution. However, this idea will allow us to generate an ensemble of random Hamiltonians without RMT eigenvalues.

We can achieve this via a Metropolis or Markov chain algorithm with some minor tweaks. First, in practice, the logarithm of $\frac{1}{\rho(\lambda(J))}$ is highly non-convex and spiky, so we will use an annealing procedure and dynamically adjust the size of the perturbations, as explained in the algorithm below. Second, it is rather difficult to control the $e^{-N\sum V(\lambda_i)}$ part of the random matrix distribution. Instead, we will just normalize our Hamiltonian so that $\tr(H^2)$ is constant. In particular, instead of dividing by the joint probability distribution, we just divide by the Vandermonde determinant. This still destroys RMU.
\newpage

The full procedure we propose is as follows:\\[6 pt]

\begin{alg}\label{metalg}
\end{alg}
\begin{enumerate}
\item Define a function $f(H, \beta_D)=-\beta_D\sum_{i<j} \ln |\lambda_i-\lambda_j|$, where $\lambda_i$ are the eigenvalues obtained by diagonalizing $H$.
\item Set $\sigma$ to some initial value (say $0.001$).
\item Run the following Metropolis algorithm at $\beta_D=0.5$ for 20000 steps:
\begin{enumerate}
    \item Sample a random new set of coefficients centered around the old ones. First we compute $l=\frac{\sigma}{2}(\sqrt{1+\frac{4x^2}{1-x^2}}-1)$ for a uniform random $x\in [0,1]$. We then sample $\binom{N}{4}$ random normal-distributed numbers in a vector, then scale the resulting vector to length $l$. Finally, we add this vector to the original coefficients. The scaling makes both smaller and larger changes more likely, and so helps deal with the spiky landscape of our potential function a bit better and reduces sensitivity to $\sigma$. Notice that $\sigma$ allows us to control the size of each step in the coupling landscape.
    \item Normalize the coefficients so that $\tr(H^2)$ is kept constant. This provides $H_{new}$.
    \item Accept the new coefficients with probability $\min(\exp(f(H_{new}, \beta_D)-f(H_{old}, \beta_D)),1)$.
    \item For every 100 steps, count the number times the coefficients were accepted. If the number is above 50, multiply $\sigma$ by 1.1; if it's less than 5, divide $\sigma$ by 1.1. The logic is that by increasing $\sigma$, we increase the size of the landscape explored with one step. This reduces the probability of finding acceptable coefficients, but allows for larger jumps. The opposite happens by reducing $\sigma$.
\end{enumerate}
\item Run the same metropolis algorithm at $\beta_D=1.0, 1.5, 2.0$ for 20000 steps each.
\end{enumerate}

Note that this procedure can get stuck at a point where $2$ or more eigenvalues are equal, due to the logarithm of their difference going to infinity. This can be dealt with by either bounding the scale $\sigma$ or simply ignoring it allowing the system to enter the spike. In either case, the algorithm appears to create non-RMT eigenvalues without changes to the OTOC or two-point functions.

\begin{figure}
    \centering
    \includegraphics[width=0.9\linewidth]{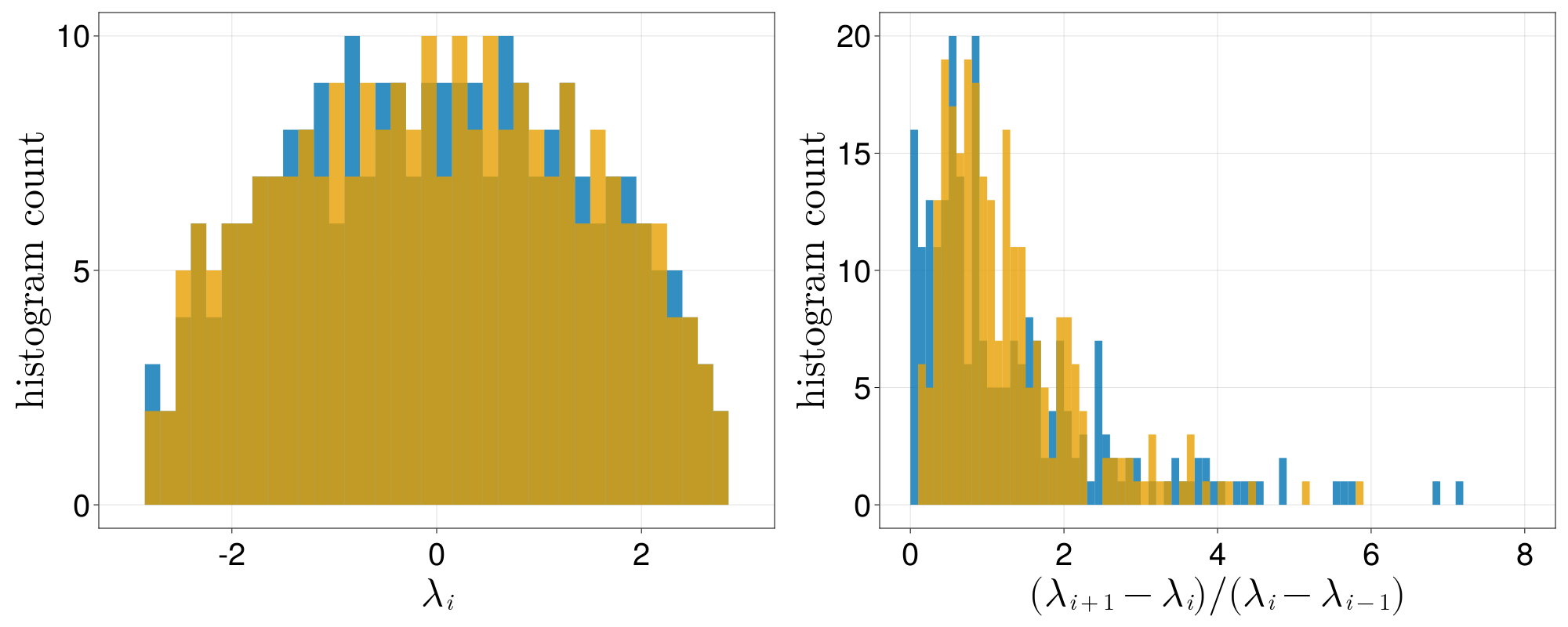}
    \caption{A histogram of the eigenvalues (left) and a histogram of the successive eigenvalue gap ratios (right), for an $N=18$ SYK model (orange) and for an $N=18$ model adjusted by the Metropolis algorithm (\ref{metalg}) (blue). The adjusted Hamiltonian is the same as the one used in Fig. \ref{fig:metrocorrelation_comp}.
    }
    \label{fig:eigapratio_metro}
\end{figure}

\begin{figure}
    \centering
    \includegraphics[width=0.9\linewidth]{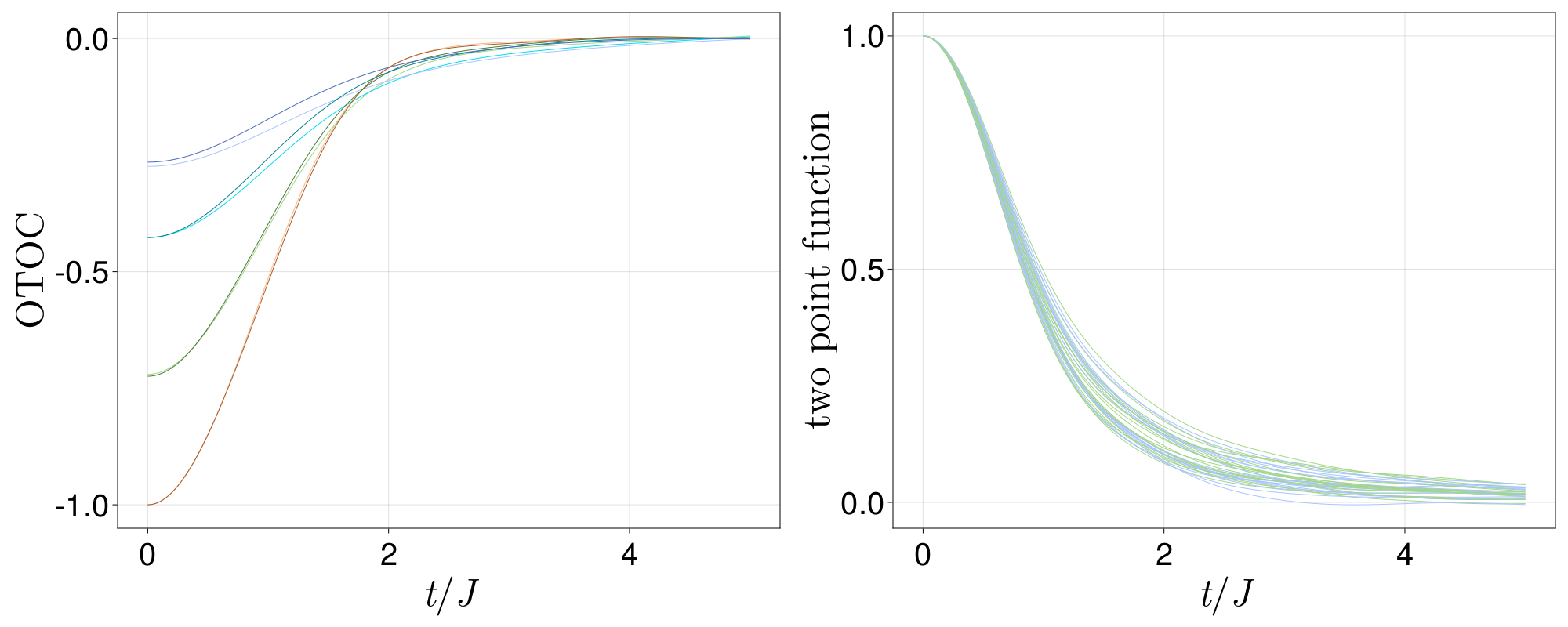}
    \caption{The OTOC (\ref{OTOCp}) for $\beta=0,1,2,3$ and the two-point functions (\ref{two}) for each of the $N=18$ fermions in the SYK model (darker colors) and the for the new Hamiltonian adjusted by the Metropolis algorithm (\ref{metalg}) (lighter colors). The adjusted Hamiltonian is the same as the one used in Fig. \ref{fig:eigapratio_metro}. 
    }
    \label{fig:metrocorrelation_comp}
\end{figure}

This algorithm results in a Hamiltonian $H_{new}$ whose eigenvalues maintain the leading density of states, but whose gap distribution is very different from the Wigner surmise associated with RMU.\footnote{We include the coefficients of this Hamiltonian in an ancillary file \texttt{coefficients.csv}.} This is depicted in Fig. (\ref{fig:eigapratio_metro}).\footnote{Albeit the difference between the blue (non-RMU) and orange (RMU) seems small in the plot, it is statistically significant, as we have observed in different draws.} This is not so surprising. After all, there are many Hamiltonians within the explored space of couplings with no spectral chaos. The real question is what has been sacrificed in the process. While we were expecting to lose something, this turned out not to be the case. The numerics show that both the density of states and correlation functions, namely
\be\label{two}
\frac{1}{Z_\beta}\text{Tr}(e^{-\beta H}O_i(t)O_i(0))
\ee
and OTOC (\ref{OTOCp}), are left unchanged in the process. This is depicted in Fig. (\ref{fig:metrocorrelation_comp}).

We conclude that, at least for these low number of spin systems, we can construct $k$-local Hamiltonians which show clear features of early-time chaos but no spectral chaos. 

\section{Discussion}\label{V}

In this article, we have revisited the Bohigas-Giannoni-Schmit (BGS) conjecture \cite{PhysRevLett.52.1}. This is also called ``the Quantum Chaos conjecture''. This conjecture is difficult to approach given the lack of a precise quantum mechanical formulation. But in recent years years it has become increasingly clear there are two types of quantum chaos: spectral chaos, controlling late-time dynamics, and eigenbasis chaos, controlling early-time dynamics such as Lyapunov growth. Each type can be approached via different quantities. The first is due to Hamiltonian spectrum statistics, while the second is due to Hamiltonian eigenvector statistics. This suggested the first working formulation of the BGS conjecture (\ref{BGS1}). Roughly, this states one cannot have early-time or eigenbasis chaos without late-time or spectral-chaos.

We then provide analytical and numerical evidence that the Poissonian ensembles developed in Ref. \cite{SpreadC,Balasubramanian:2023kwd} provide families of counterexamples to the conjecture in such a first formulation. Moreover, these counterexamples can be constructed for any example satisfying the conjecture. This stems from the basic fact that eigenvalues and eigenvectors are, a priori, uncorrelated. However, the large zoo of examples verifying the BGS conjecture suggests the first formulation lacks some assumption. A most natural one is $k$-locality. This is a realistic/physical way of constraining the type of Hamiltonians of interest. It relates in intricate manners eigenvector and eigenvalue statistics. As could be anticipated, we then demonstrated that the Poissonian ensembles are not $k$-local. Adding the assumption of Hamiltonian $k$-locality leads to the second formulation of the BGS conjecture (\ref{BGS2}). This avoids the otherwise insightful Poissonian counterexamples.

In the last section (\ref{IV}) we tested the more accurate second formulation. This was accomplished by exploring the space of $4$-local SYK-like Hamiltonians via a Metropolis or Markov-type algorithm. The idea behind the algorithm is to reward directions in coupling space in which we erase spectral statistics. Contrary to expectations, for the case of $18$-Majorana fermions, we find Hamiltonians that show features of eigenbasis chaos but not spectral chaos.

We now discuss possible wayouts, as well as applications of these intriguing results in the context of quantum gravity.

\subsection{On the thermodynamic limit of the Metropolis algorithm}

The idea behind the construction of Hamiltonians that violate the second formulation of the BGS conjecture was simple. Instead of taking Gaussian random couplings in SYK, we look for distributions with similar properties but assign a high probability to eigenvalue degeneracies. Unfortunately, we could only work out the case of $18$ spins. The case of $22$ spins is also doable and we will explore it in detail in future work.

But in both cases, there is something undesirable that may save the conjecture. First, the constructed Hamiltonian is not a proper counterexample to the conjecture. The reason is that for $18$ spins we cannot claim the result is valid in the thermodynamic limit. The objective for future work is to apply these ideas to systems with a larger number of spins, eventually taking this number to infinity.

Notice that for certain problems, e.g. aspects of quantum thermalization that depend on Hilbert space dimension, a low number of spins suffice. But here we encounter a counting issue. For a low number of spins, in particular, for $18$ and $22$ Majorana fermions, the number of independent couplings is of the order of the number of Hamiltonian eigenvalues. The first grows as $N^4$, while the second as $2^{N/2}$. For large-$N$, the number of eigenvalues is exceedingly larger than the number of couplings. But at small $N$ this is not the case. A possibility is that in the large $N$ limit, modifications in the ``small'' number of couplings will not be able to erase spectral chaos without erasing eigenbasis chaos. Elucidating this issue is an important problem for future research.

\subsection{BGS in Quantum Field Theory}

If we could claim the results of section (\ref{IV}) are also valid in the thermodynamic limit, we would conclude that the second version of the BGS conjecture is also invalid. In this scenario, the next logical possibility to constrain the space of Hamiltonians is to add spatial locality (apart from $k$-locality). This naturally leads to a formulation of the conjecture in QFT as

\begin{conj}
    {\sl A local quantum field theory whose correlation functions thermalize and OTOC shows Lyapunov growth, i.e. displaying eigenbasis chaos, also displays spectral chaos. }
\end{conj}

The idea is that QFT seems the best scenario to include spatial and $k$-locality into a tight mathematical structure. Indeed, while locality is a foundational principle in QFT, we find that $k$-locality is also implicitly included. In free or weakly coupled theories with a Lagrangian formulation, this is obvious. More abstractly, the stress tensor in a CFT has a fixed dimension $d$. Higher dimension operators are irrelevant for the dynamics at long distances. The notion of scaling dimension is the natural analog of $k$-locality in QFT, matching well with the notion of $k$-locality in appropriate limits or lattice formulations.\footnote{For a similar discussion in the context of circuit complexity in QFT see \cite{Magan:2018nmu}.} It is interesting to explore whether changes in the spectrum of the sort explored in this article can be done without spoiling space-time locality.

\subsection{Quantum Gravity}

Black holes are maximally chaotic, in the precise sense of saturating the bound on chaos \cite{Maldacena:2015waa}. From the present perspective, they display early-time chaos.\footnote{A simple path to this last conclusion uses the hyperbolic optical metric \cite{Barbon:2011pn,Barbon:2012zv}.} Following the BGS conjecture, we expect them to display spectral chaos as well. Signatures of spectral chaos in black hole physics and holography have attracted a lot of attention recently, leading to important breakthroughs \cite{Cotler:2016fpe,Saad:2018bqo,Saad:2019lba,Stanford:2019vob,stanford2020jt,Cotler:2020ugk,DiUbaldo:2023qli,Belin:2023efa}. In these contributions, it has been shown how wormhole configurations in the gravitational path integral codify the universal correlations of RMT. For example, the ``double cone'' geometry described in \cite{Saad:2018bqo} provides the ramp in the spectral form factor of the theory, a smoking gun for spectral chaos.

In the light of our results, we wish to make some remarks in this context. If we were able to extend the Metropolis construction to a large number of fermions, we would obtain a toy model of a quantum theory of gravity without spectral chaos. This would be a model with the same behavior as SYK, e.g. density of states and correlation functions, but with different rules for the non-perturbative corrections. In particular, the non-perturbative corrections to spectral statistics would be fully controlled by those found in section (\ref{II}). This model would even be $k$-local.

But in the quantum gravity context, it is not clear whether we should add the assumption of $k$-locality to the \emph{boundary theory}. Let us explain. Consider a generic AdS/CFT scenario. This could be a low-dimension model \cite{Mertens:2022irh}, e.g. JT gravity or SYK, or a full-fledged top-down model in higher dimensions, such as those appearing in String Theory \cite{Aharony:1999ti}. For any such scenario, we can define the associated Poissonian ensembles of Ref. \cite{SpreadC,Balasubramanian:2023kwd}. For SYK, as we have seen, this leads to a non $k$-local Hamiltonian. For higher dimensional examples, this would break the locality of the CFT at a fine-grained level. But it is unclear to us whether these are important/fundamental/necessary properties from a bulk perspective. Equivalently, the effective field theory in the bulk would be the same for all orders in a perturbative expansion in Newton's constant. Differences would only show up in the definition of the theory at the non-perturbative level. These non-perturbative corrections destroy the notion of bulk locality in both cases, whether we non-perturbatively define the theory with RMU or Poissonian statistics.

These observations might be better explored using algebraic methods for quantum gravity, following \cite{Leutheusser:2021frk,Witten:2021unn,Chandrasekaran:2022cip,Witten:2023qsv,Leutheusser:2024yvf}. The question there would be, do the type III or type II emergent algebras in the thermodynamic limit depend on fine-grained details of the spectrum of the Hamiltonian. We are inclined to think that the emergent algebras would see no difference, but it is an interesting avenue for future exploration.

One possible objection is that part of the lessons/utility of the non-perturbative corrections discovered recently, captured by the gravitational path integral, was to solve fundamental issues in black hole physics. Most notably the information paradox in all its different manifestations, e.g. unitarity of black hole evaporation \cite{Penington:2019npb,Almheiri:2019psf,Almheiri:2019qdq,Penington:2019kki}, finiteness of black hole entropy \cite{Balasubramanian:2022gmo} and long time saturation of correlation functions \cite{Maldacena:2001kr,Cotler:2016fpe}. But the Poissonian ensembles are unitary theories with the right entropy, so all those problems should disappear as well. We discuss some of these aspects in more detail in the next section.

Therefore, as far as we can see at present, the Poissonian Hamiltonian ensembles seem potential candidates for theories of quantum gravity. As discussed in section (\ref{II}), they are even typical if we apply the Jaynes maximum ignorance principle and just fix the Bekenstein-Haking density of states. In turn, this is the less biased choice from a bulk perspective. These theories solve in a simple manner several key issues in quantum gravity. But they accomplish so through a drastically different set of rules for the gravitational path integral. More concretely, for these theories, one should blindly apply the spectral statistics found in section (\ref{II}) and their consequences in products of partition functions.

\subsection{Black hole Microstates from Thermofield Doubles}

Another application of the present ideas to the black hole and quantum gravity context goes as follows. The problem of constructing finite-dimensional Hilbert spaces in quantum gravity has seen a huge development in recent years \cite{Penington:2019kki,Hsin:2020mfa,Chandra:2022fwi,Balasubramanian:2022gmo,Balasubramanian:2022lnw,Boruch:2023trc,Antonini:2023hdh,Climent:2024trz,Iliesiu:2024cnh,Boruch:2024kvv}. These constructions use various sorts of heavy (backreacting) operators. There is yet another family of black hole microstates which makes transparent the connection with different avatars of the information paradox. This is the family of time-evolving thermofield doubles (TFD) \cite{Papadodimas:2015xma,Banerjee:2023liw}. Starting from the usual TFD
\be 
\vert\psi_{\beta}\rangle \equiv\frac{1}{\sqrt{Z_{\beta}}}\sum_n e^{-\frac{\beta E_n}{2}}\vert n,n\rangle\;,
\label{TFDdef}
\ee
unitary evolution with a single Hamiltonian gives
\be 
\vert\psi_{\beta} (t)\rangle=e^{-iHt}\vert\psi_{\beta}\rangle=\vert\psi_{\beta+2it}\rangle=\sum_n e^{-(\beta+2it) E_n/2}\,\vert n,n\rangle\;.
\ee
This can be understood as a continuous family of black hole microstates (labeled by the time) for the eternal black hole geometry \cite{Maldacena:2001kr}. In particular, all of them are exactly thermal when observed from one side.

Following the discussion on the previous item, we now assume a Poissonian non-perturbative completion of quantum gravity. Then it is simple to prove these time-evolved TFD states form a basis of the black hole Hilbert space. We follow the usual ideology. We start by taking a discrete set of $\Omega$ microstates, in this case by taking a discrete set of times $t_i$ and associated $\vert \psi_{\beta} (t_j)\rangle$. We assume $t_1$ is sufficiently large and that $t_j=j t_1$, with $j=1,\cdots , \Omega$. This way all time differences $t_i-t_j$ are large. To compute the dimension of the Hilbert space spanned by this discrete set of black hole microstates, we compute the rank of the matrix of overlaps. This matrix is defined here as
\be 
G_{i j}\equiv \langle \psi_{\beta} (t_i)\vert \psi_{\beta} (t_j)\rangle\;,
\ee
where $i,j=1,\cdots, \Omega$. We can compute precise statistical features of this matrix using the Poissonian ensemble, in particular the statistics found in section (\ref{II}). The first moment is equal to the inner product between microstates. This is the analytically continued partition function $Z_{\beta-i\Delta t_{jk}}$, where $\Delta t_{jk}\equiv t_j-t_k $ is the difference in times defining the different TFD states. On average in the Poissonian ensemble, this is
\be 
\overline{Z_{\beta-i\Delta t_{jk}}}=\int dE \overline{\rho(E)} e^{-(\beta-i\Delta t_{jk}) E}\rightarrow 0\;,
\ee
since the time difference can be taken as large as we want. The fact the first moment is zero does not mean all others are zero as well \cite{Penington:2019kki}. Consider the $n$-moment
\be 
G_{1 2}G_{2 3}\cdots G_{\Omega 1}=\langle \psi_{\beta} (t_1)\vert \psi_{\beta} (t_2)\rangle\langle \psi_{\beta} (t_2)\vert \psi_{\beta} (t_3)\rangle\cdots \langle \psi_{\beta} (t_n)\vert \psi_{\beta} (t_1)\rangle\;.
\ee
In the energy basis, it reads
\be 
G_{1 2}\cdots G_{\Omega 1}= \frac{1}{Z_{\beta}^n}\sum_{j_1,\cdots j_n}\,e^{-\beta\sum_{l=1}^{n}E_{j_l}-i t_l (E_{j_{l+1}}-E_{j_l})}\;.
\ee
The connected contribution to the $n$-moment follows from the connected contribution to the $n$-moment of the density of states. This was found to be
\be 
\overline{\rho(E_1)\cdots\rho(E_n)}\vert_{c}=\frac{1}{N^{n-1}} \overline{\rho(E_1)}\prod_{j=1}^n \,\delta (E_1-E_j)\;.
\ee
The connected $n$-moment of the Gram matrix of overlaps then reads
\be 
\overline{G_{1 2}G_{2 3}\cdots G_{\Omega 1}}\vert_{c}=\frac{Z_{n\beta}}{Z_{\beta}^n}\;.
\ee
The statistics $\frac{Z(n\beta)}{Z_{\beta}^n}$ are the universal statistics found in \cite{Balasubramanian:2022gmo,Balasubramanian:2022lnw,Climent:2024trz} for black hole microstates constructed with heavy shells. From such statistics one derives that the dimension of the microcanonical Hilbert space spanned by $\vert \psi_{\beta} (t_j)\rangle$ with $j=1,\cdots , \Omega$ is equal to $\Omega$ if $\Omega$ is smaller than the microcanonical degeneracy $e^{S(E)}$, and it is equal to the microcanonical degeneracy $e^{S(E)}$ if $\Omega$ is larger than $e^{S(E)}$.

Let us note that for $n=2$ the statistics of the Gram matrix of these black hole microstates is equal to the plateau of the spectral form factor, related to Maldacena's avatar of the information paradox \cite{Maldacena:2001kr,Cotler:2016fpe}.\footnote{From a gravitational perspective, one could still desire to find a two boundary wormhole providing these Poissonian statistics, i.e. the plateau.} Besides, it is transparent from this standpoint that the problem of deriving the saturation of the plateau of the spectral form factor is that of proving the time-evolved TFD states form a basis of the Hilbert space. For the Poissonian ensembles, access to the plateau is simple from the ensemble perspective,  see ec. (\ref{asff}), allowing the derivation. For theories with spectral chaos, it should be possible as well, but the path is obstructed for technical reasons. See \cite{Blommaert:2022lbh,Saad:2022kfe} for recent developments in low dimensional models.

\medskip
{\bf Acknowledgements.} 

 We are grateful to Vijay Balasubramanian, Horacio Casini, Tom Hartman, Eric Perlmutter, Martin Sasieta and Steve Shenker for useful discussions. We also wish to thank the participants of the workshop ``The microscopic origin of black hole entropy'', held in the Aspen Centre for Physics (ACP). This work was performed in part at the ACP, which is supported by a grant from the Simons Foundation (1161654, Troyer).  The work of JM is supported by CONICET, Argentina. J.M acknowledges hospitality and support from the International Institute of Physics, Natal, through Simons Foundation award number 1023171-RC. The work of QW is supported by a DOE through DE-SC0013528,  QuantISED grant DE-SC0020360, and the Simons Foundation It From Qubit collaboration (385592).

\bibliographystyle{utphys}
\bibliography{main}

\providecommand{\href}[2]{#2}\begingroup\raggedright\begin{thebibliography}{10}

\bibitem{10.2307/1969956}
E.~P. Wigner, ``Characteristics vectors of bordered matrices with infinite
  dimensions ii,'' {\em Annals of Mathematics} {\bfseries 65} no.~2, (1957)
  203--207. \url{http://www.jstor.org/stable/1969956}.

\bibitem{osti_4801180}
F.~J. Dyson, ``Statistical theory of the energy levels of complex systems. i,''
  \href{http://dx.doi.org/10.1063/1.1703773}{{\em Journal of Mathematical
  Physics} {\bfseries 3} no.~1, (1, 1962) }.
  \url{https://www.osti.gov/biblio/4801180}.

\bibitem{PhysRevLett.52.1}
O.~Bohigas, M.~J. Giannoni, and C.~Schmit, ``Characterization of chaotic
  quantum spectra and universality of level fluctuation laws,''
  \href{http://dx.doi.org/10.1103/PhysRevLett.52.1}{{\em Phys. Rev. Lett.}
  {\bfseries 52} (Jan, 1984) 1--4}.
  \url{https://link.aps.org/doi/10.1103/PhysRevLett.52.1}.

\bibitem{bookhaake}
F.~Haake, \href{http://dx.doi.org/10.1007/978-3-662-04506-0}{{\em Quantum
  Signatures of Chaos}}, vol.~54.
\newblock 01, 2001.

\bibitem{Guhr_1998}
T.~Guhr, A.~Müller–Groeling, and H.~A. Weidenmüller, ``Random-matrix
  theories in quantum physics: common concepts,''
  \href{http://dx.doi.org/10.1016/s0370-1573(97)00088-4}{{\em Physics Reports}
  {\bfseries 299} no.~4–6, (June, 1998) 189–425}.
  \url{http://dx.doi.org/10.1016/S0370-1573(97)00088-4}.

\bibitem{akemann2011oxford}
G.~Akemann, J.~Baik, and P.~Di~Francesco, {\em The Oxford Handbook of Random
  Matrix Theory}.
\newblock Oxford Handbooks in Mathematics. OUP Oxford, 2011.
\newblock \url{https://books.google.com/books?id=QhsgKQEACAAJ}.

\bibitem{Andreev:1996tw}
A.~V. Andreev, O.~Agam, B.~L. Altshuler, and B.~D. Simons, ``{Quantum chaos,
  irreversible classical dynamics, and random matrix theory},''
  \href{http://dx.doi.org/10.1103/PhysRevLett.76.3947}{{\em Phys. Rev. Lett.}
  {\bfseries 76} (1996) 3947--3950},
  \href{http://arxiv.org/abs/cond-mat/9601001}{{\ttfamily
  arXiv:cond-mat/9601001}}.

\bibitem{Andreev:1996twa}
A.~V. Andreev, B.~D. Simons, O.~Agam, and B.~L. Altshuler, ``{Semiclassical
  field theory approach to quantum chaos},''
  \href{http://dx.doi.org/10.1016/S0550-3213(96)00473-7}{{\em Nucl. Phys. B}
  {\bfseries 482} (1996) 536--566},
  \href{http://arxiv.org/abs/cond-mat/9605204}{{\ttfamily
  arXiv:cond-mat/9605204}}.

\bibitem{Altland:2014wna}
A.~Altland, S.~Gnutzmann, F.~Haake, and T.~Micklitz, ``{A review of sigma
  models for quantum chaotic dynamics},''
  \href{http://dx.doi.org/10.1088/0034-4885/78/8/086001}{{\em Rept. Prog.
  Phys.} {\bfseries 78} no.~8, (2015) 086001},
  \href{http://arxiv.org/abs/1412.5336}{{\ttfamily arXiv:1412.5336 [nlin.CD]}}.

\bibitem{Gutzwiller}
M.~C. {Gutzwiller}, ``{Periodic Orbits and Classical Quantization
  Conditions},'' \href{http://dx.doi.org/10.1063/1.1665596}{{\em Journal of
  Mathematical Physics} {\bfseries 12} no.~3, (Mar., 1971) 343--358}.

\bibitem{1985RSPSA.400..229B}
M.~V. {Berry}, ``{Semiclassical Theory of Spectral Rigidity},''
  \href{http://dx.doi.org/10.1098/rspa.1985.0078}{{\em Proceedings of the Royal
  Society of London Series A} {\bfseries 400} no.~1819, (Aug., 1985) 229--251}.

\bibitem{jh}
J.~H. Hannay and A.~M. O.~D. Almeida, ``Periodic orbits and a correlation
  function for the semiclassical density of states,''
  \href{http://dx.doi.org/10.1088/0305-4470/17/18/013}{{\em Journal of Physics
  A: Mathematical and General} {\bfseries 17} no.~18, (Dec, 1984) 3429}.
  \url{https://dx.doi.org/10.1088/0305-4470/17/18/013}.

\bibitem{Larkin1969QuasiclassicalMI}
A.~I. Larkin and Y.~N. Ovchinnikov, ``Quasiclassical method in the theory of
  superconductivity,'' {\em Journal of Experimental and Theoretical Physics}
  (1969) . \url{https://api.semanticscholar.org/CorpusID:117608877}.

\bibitem{Maldacena:2015waa}
J.~Maldacena, S.~H. Shenker, and D.~Stanford, ``{A bound on chaos},''
  \href{http://dx.doi.org/10.1007/JHEP08(2016)106}{{\em JHEP} {\bfseries 08}
  (2016) 106}, \href{http://arxiv.org/abs/1503.01409}{{\ttfamily
  arXiv:1503.01409 [hep-th]}}.

\bibitem{PhysRevA.43.2046}
J.~M. Deutsch, ``Quantum statistical mechanics in a closed system,''
  \href{http://dx.doi.org/10.1103/PhysRevA.43.2046}{{\em Phys. Rev. A}
  {\bfseries 43} (Feb, 1991) 2046--2049}.
  \url{https://link.aps.org/doi/10.1103/PhysRevA.43.2046}.

\bibitem{PhysRevE.50.888}
M.~Srednicki, ``Chaos and quantum thermalization,''
  \href{http://dx.doi.org/10.1103/PhysRevE.50.888}{{\em Phys. Rev. E}
  {\bfseries 50} (Aug, 1994) 888--901}.
  \url{https://link.aps.org/doi/10.1103/PhysRevE.50.888}.

\bibitem{PhysRevE.99.042139}
L.~Foini and J.~Kurchan, ``Eigenstate thermalization hypothesis and out of time
  order correlators,'' \href{http://dx.doi.org/10.1103/PhysRevE.99.042139}{{\em
  Phys. Rev. E} {\bfseries 99} (Apr, 2019) 042139}.
  \url{https://link.aps.org/doi/10.1103/PhysRevE.99.042139}.

\bibitem{SpreadC}
V.~Balasubramanian, P.~Caputa, J.~M. Magan, and Q.~Wu, ``{Quantum chaos and the
  complexity of spread of states},''
  \href{http://dx.doi.org/10.1103/PhysRevD.106.046007}{{\em Phys. Rev. D}
  {\bfseries 106} no.~4, (2022) 046007},
  \href{http://arxiv.org/abs/2202.06957}{{\ttfamily arXiv:2202.06957
  [hep-th]}}.

\bibitem{Balasubramanian:2023kwd}
V.~Balasubramanian, J.~M. Magan, and Q.~Wu, ``{Quantum chaos, integrability,
  and late times in the Krylov basis},''
  \href{http://arxiv.org/abs/2312.03848}{{\ttfamily arXiv:2312.03848
  [hep-th]}}.

\bibitem{kitaev}
A.~Kitaev, ``{A simple model of quantum holography},'' {\em Talks at KITP,
  April 7, 2015 and May 27, 2015} .

\bibitem{sachdev}
S.~Sachdev and J.~Ye, ``{Gapless spin fluid ground state in a random, quantum
  Heisenberg ferromagnet},'' {\em Phys. Rev. Lett.} {\bfseries 70} (1993) 3339,
  \href{http://arxiv.org/abs/9212030}{{\ttfamily arXiv:9212030 [cond-mat]}}.

\bibitem{Erdmenger:2023shk}
J.~Erdmenger, S.-K. Jian, and Z.-Y. Xian, ``{Universal chaotic dynamics from
  Krylov space},'' \href{http://arxiv.org/abs/2303.12151}{{\ttfamily
  arXiv:2303.12151 [hep-th]}}.

\bibitem{Nandy:2024zcd}
P.~Nandy, ``{Tridiagonal Hamiltonians modeling the density of states of the
  Double-Scaled SYK model},'' \href{http://arxiv.org/abs/2410.07847}{{\ttfamily
  arXiv:2410.07847 [hep-th]}}.

\bibitem{Gu:2024hmj}
A.~Gu, Y.~Quek, S.~Yelin, J.~Eisert, and L.~Leone, ``{Simulating quantum chaos
  without chaos},'' \href{http://arxiv.org/abs/2410.18196}{{\ttfamily
  arXiv:2410.18196 [quant-ph]}}.

\bibitem{Lee:2024zvj}
W.~Lee, H.~Kwon, and G.~Y. Cho, ``{Pseudochaotic Many-Body Dynamics as a
  Pseudorandom State Generator},''
  \href{http://arxiv.org/abs/2410.21268}{{\ttfamily arXiv:2410.21268
  [quant-ph]}}.

\bibitem{dssyk}
V.~Balasubramanian, J.~M. Magan, P.~Nandi, and Q.~Wu, ``{Hilbert space
  construction, ER bridges, and spread complexity in the Double Scaled SYK
  Model},'' \href{http://arxiv.org/abs/To appear soon}{{\ttfamily To appear
  soon}}.

\bibitem{Lin:2022zxd}
H.~W. Lin, J.~Maldacena, L.~Rozenberg, and J.~Shan, ``{Looking at
  supersymmetric black holes for a very long time},''
  \href{http://dx.doi.org/10.21468/SciPostPhys.14.5.128}{{\em SciPost Phys.}
  {\bfseries 14} no.~5, (2023) 128},
  \href{http://arxiv.org/abs/2207.00408}{{\ttfamily arXiv:2207.00408
  [hep-th]}}.

\bibitem{Chen:2024oqv}
Y.~Chen, H.~W. Lin, and S.~H. Shenker, ``{BPS Chaos},''
  \href{http://arxiv.org/abs/2407.19387}{{\ttfamily arXiv:2407.19387
  [hep-th]}}.

\bibitem{Cotler:2016fpe}
J.~S. Cotler, G.~Gur-Ari, M.~Hanada, J.~Polchinski, P.~Saad, S.~H. Shenker,
  D.~Stanford, A.~Streicher, and M.~Tezuka, ``{Black Holes and Random
  Matrices},'' \href{http://dx.doi.org/10.1007/JHEP05(2017)118}{{\em JHEP}
  {\bfseries 05} (2017) 118}, \href{http://arxiv.org/abs/1611.04650}{{\ttfamily
  arXiv:1611.04650 [hep-th]}}. [Erratum: JHEP 09, 002 (2018)].

\bibitem{Maldacena:2001kr}
J.~M. Maldacena, ``{Eternal black holes in anti-de Sitter},''
  \href{http://dx.doi.org/10.1088/1126-6708/2003/04/021}{{\em JHEP} {\bfseries
  04} (2003) 021}, \href{http://arxiv.org/abs/hep-th/0106112}{{\ttfamily
  arXiv:hep-th/0106112}}.

\bibitem{Jafferis:2022wez}
D.~L. Jafferis, D.~K. Kolchmeyer, B.~Mukhametzhanov, and J.~Sonner,
  ``{Jackiw-Teitelboim gravity with matter, generalized eigenstate
  thermalization hypothesis, and random matrices},''
  \href{http://dx.doi.org/10.1103/PhysRevD.108.066015}{{\em Phys. Rev. D}
  {\bfseries 108} no.~6, (2023) 066015},
  \href{http://arxiv.org/abs/2209.02131}{{\ttfamily arXiv:2209.02131
  [hep-th]}}.

\bibitem{You:2016ldz}
Y.-Z. You, A.~W.~W. Ludwig, and C.~Xu, ``{Sachdev-Ye-Kitaev Model and
  Thermalization on the Boundary of Many-Body Localized Fermionic Symmetry
  Protected Topological States},''
  \href{http://dx.doi.org/10.1103/PhysRevB.95.115150}{{\em Phys. Rev. B}
  {\bfseries 95} no.~11, (2017) 115150},
  \href{http://arxiv.org/abs/1602.06964}{{\ttfamily arXiv:1602.06964
  [cond-mat.str-el]}}.

\bibitem{PhysRev.106.620}
E.~T. Jaynes, ``Information theory and statistical mechanics,''
  \href{http://dx.doi.org/10.1103/PhysRev.106.620}{{\em Phys. Rev.} {\bfseries
  106} (May, 1957) 620--630}.
  \url{https://link.aps.org/doi/10.1103/PhysRev.106.620}.

\bibitem{Magan:2018nmu}
J.~M. Mag\'an, ``{Black holes, complexity and quantum chaos},''
  \href{http://dx.doi.org/10.1007/JHEP09(2018)043}{{\em JHEP} {\bfseries 09}
  (2018) 043}, \href{http://arxiv.org/abs/1805.05839}{{\ttfamily
  arXiv:1805.05839 [hep-th]}}.

\bibitem{Barbon:2011pn}
J.~L.~F. Barbon and J.~M. Magan, ``{Chaotic Fast Scrambling At Black Holes},''
  \href{http://dx.doi.org/10.1103/PhysRevD.84.106012}{{\em Phys. Rev. D}
  {\bfseries 84} (2011) 106012},
  \href{http://arxiv.org/abs/1105.2581}{{\ttfamily arXiv:1105.2581 [hep-th]}}.

\bibitem{Barbon:2012zv}
J.~L.~F. Barbon and J.~M. Magan, ``{Fast Scramblers, Horizons and Expander
  Graphs},'' \href{http://dx.doi.org/10.1007/JHEP08(2012)016}{{\em JHEP}
  {\bfseries 08} (2012) 016}, \href{http://arxiv.org/abs/1204.6435}{{\ttfamily
  arXiv:1204.6435 [hep-th]}}.

\bibitem{Saad:2018bqo}
P.~Saad, S.~H. Shenker, and D.~Stanford, ``{A semiclassical ramp in SYK and in
  gravity},'' \href{http://arxiv.org/abs/1806.06840}{{\ttfamily
  arXiv:1806.06840 [hep-th]}}.

\bibitem{Saad:2019lba}
P.~Saad, S.~H. Shenker, and D.~Stanford, ``{JT gravity as a matrix integral},''
  \href{http://arxiv.org/abs/1903.11115}{{\ttfamily arXiv:1903.11115
  [hep-th]}}.

\bibitem{Stanford:2019vob}
D.~Stanford and E.~Witten, ``{JT gravity and the ensembles of random matrix
  theory},'' \href{http://dx.doi.org/10.4310/ATMP.2020.v24.n6.a4}{{\em Adv.
  Theor. Math. Phys.} {\bfseries 24} no.~6, (2020) 1475--1680},
  \href{http://arxiv.org/abs/1907.03363}{{\ttfamily arXiv:1907.03363
  [hep-th]}}.

\bibitem{stanford2020jt}
D.~Stanford and E.~Witten, ``Jt gravity and the ensembles of random matrix
  theory,'' 2020.

\bibitem{Cotler:2020ugk}
J.~Cotler and K.~Jensen, ``{AdS$_{3}$ gravity and random CFT},''
  \href{http://dx.doi.org/10.1007/JHEP04(2021)033}{{\em JHEP} {\bfseries 04}
  (2021) 033}, \href{http://arxiv.org/abs/2006.08648}{{\ttfamily
  arXiv:2006.08648 [hep-th]}}.

\bibitem{DiUbaldo:2023qli}
G.~Di~Ubaldo and E.~Perlmutter, ``{AdS$_{3}$/RMT$_{2}$ duality},''
  \href{http://dx.doi.org/10.1007/JHEP12(2023)179}{{\em JHEP} {\bfseries 12}
  (2023) 179}, \href{http://arxiv.org/abs/2307.03707}{{\ttfamily
  arXiv:2307.03707 [hep-th]}}.

\bibitem{Belin:2023efa}
A.~Belin, J.~de~Boer, D.~L. Jafferis, P.~Nayak, and J.~Sonner, ``{Approximate
  CFTs and Random Tensor Models},''
  \href{http://arxiv.org/abs/2308.03829}{{\ttfamily arXiv:2308.03829
  [hep-th]}}.

\bibitem{Mertens:2022irh}
T.~G. Mertens and G.~J. Turiaci, ``{Solvable models of quantum black holes: a
  review on Jackiw\textendash{}Teitelboim gravity},''
  \href{http://dx.doi.org/10.1007/s41114-023-00046-1}{{\em Living Rev. Rel.}
  {\bfseries 26} no.~1, (2023) 4},
  \href{http://arxiv.org/abs/2210.10846}{{\ttfamily arXiv:2210.10846
  [hep-th]}}.

\bibitem{Aharony:1999ti}
O.~Aharony, S.~S. Gubser, J.~M. Maldacena, H.~Ooguri, and Y.~Oz, ``{Large N
  field theories, string theory and gravity},''
  \href{http://dx.doi.org/10.1016/S0370-1573(99)00083-6}{{\em Phys. Rept.}
  {\bfseries 323} (2000) 183--386},
  \href{http://arxiv.org/abs/hep-th/9905111}{{\ttfamily arXiv:hep-th/9905111}}.

\bibitem{Leutheusser:2021frk}
S.~A.~W. Leutheusser and H.~Liu, ``{Emergent Times in Holographic Duality},''
  \href{http://dx.doi.org/10.1103/PhysRevD.108.086020}{{\em Phys. Rev. D}
  {\bfseries 108} no.~8, (2023) 086020},
  \href{http://arxiv.org/abs/2112.12156}{{\ttfamily arXiv:2112.12156
  [hep-th]}}.

\bibitem{Witten:2021unn}
E.~Witten, ``{Gravity and the crossed product},''
  \href{http://dx.doi.org/10.1007/JHEP10(2022)008}{{\em JHEP} {\bfseries 10}
  (2022) 008}, \href{http://arxiv.org/abs/2112.12828}{{\ttfamily
  arXiv:2112.12828 [hep-th]}}.

\bibitem{Chandrasekaran:2022cip}
V.~Chandrasekaran, R.~Longo, G.~Penington, and E.~Witten, ``{An algebra of
  observables for de Sitter space},''
  \href{http://dx.doi.org/10.1007/JHEP02(2023)082}{{\em JHEP} {\bfseries 02}
  (2023) 082}, \href{http://arxiv.org/abs/2206.10780}{{\ttfamily
  arXiv:2206.10780 [hep-th]}}.

\bibitem{Witten:2023qsv}
E.~Witten, ``{Algebras, regions, and observers.},''
  \href{http://dx.doi.org/10.1090/pspum/107/01954}{{\em Proc. Symp. Pure Math.}
  {\bfseries 107} (2024) 247--276},
  \href{http://arxiv.org/abs/2303.02837}{{\ttfamily arXiv:2303.02837
  [hep-th]}}.

\bibitem{Leutheusser:2024yvf}
S.~Leutheusser and H.~Liu, ``{Superadditivity in large $N$ field theories and
  performance of quantum tasks},''
  \href{http://arxiv.org/abs/2411.04183}{{\ttfamily arXiv:2411.04183
  [hep-th]}}.

\bibitem{Penington:2019npb}
G.~Penington, ``{Entanglement Wedge Reconstruction and the Information
  Paradox},'' \href{http://dx.doi.org/10.1007/JHEP09(2020)002}{{\em JHEP}
  {\bfseries 09} (2020) 002}, \href{http://arxiv.org/abs/1905.08255}{{\ttfamily
  arXiv:1905.08255 [hep-th]}}.

\bibitem{Almheiri:2019psf}
A.~Almheiri, N.~Engelhardt, D.~Marolf, and H.~Maxfield, ``{The entropy of bulk
  quantum fields and the entanglement wedge of an evaporating black hole},''
  \href{http://dx.doi.org/10.1007/JHEP12(2019)063}{{\em JHEP} {\bfseries 12}
  (2019) 063}, \href{http://arxiv.org/abs/1905.08762}{{\ttfamily
  arXiv:1905.08762 [hep-th]}}.

\bibitem{Almheiri:2019qdq}
A.~Almheiri, T.~Hartman, J.~Maldacena, E.~Shaghoulian, and A.~Tajdini,
  ``{Replica Wormholes and the Entropy of Hawking Radiation},''
  \href{http://dx.doi.org/10.1007/JHEP05(2020)013}{{\em JHEP} {\bfseries 05}
  (2020) 013}, \href{http://arxiv.org/abs/1911.12333}{{\ttfamily
  arXiv:1911.12333 [hep-th]}}.

\bibitem{Penington:2019kki}
G.~Penington, S.~H. Shenker, D.~Stanford, and Z.~Yang, ``{Replica wormholes and
  the black hole interior},''
  \href{http://dx.doi.org/10.1007/JHEP03(2022)205}{{\em JHEP} {\bfseries 03}
  (2022) 205}, \href{http://arxiv.org/abs/1911.11977}{{\ttfamily
  arXiv:1911.11977 [hep-th]}}.

\bibitem{Balasubramanian:2022gmo}
V.~Balasubramanian, A.~Lawrence, J.~M. Magan, and M.~Sasieta, ``{Microscopic
  Origin of the Entropy of Black Holes in General Relativity},''
  \href{http://dx.doi.org/10.1103/PhysRevX.14.011024}{{\em Phys. Rev. X}
  {\bfseries 14} no.~1, (2024) 011024},
  \href{http://arxiv.org/abs/2212.02447}{{\ttfamily arXiv:2212.02447
  [hep-th]}}.

\bibitem{Hsin:2020mfa}
P.-S. Hsin, L.~V. Iliesiu, and Z.~Yang, ``{A violation of global symmetries
  from replica wormholes and the fate of black hole remnants},''
  \href{http://dx.doi.org/10.1088/1361-6382/ac2134}{{\em Class. Quant. Grav.}
  {\bfseries 38} no.~19, (2021) 194004},
  \href{http://arxiv.org/abs/2011.09444}{{\ttfamily arXiv:2011.09444
  [hep-th]}}.

\bibitem{Chandra:2022fwi}
J.~Chandra and T.~Hartman, ``{Coarse graining pure states in AdS/CFT},''
  \href{http://dx.doi.org/10.1007/JHEP10(2023)030}{{\em JHEP} {\bfseries 10}
  (2023) 030}, \href{http://arxiv.org/abs/2206.03414}{{\ttfamily
  arXiv:2206.03414 [hep-th]}}.

\bibitem{Balasubramanian:2022lnw}
V.~Balasubramanian, A.~Lawrence, J.~M. Magan, and M.~Sasieta, ``{Microscopic
  Origin of the Entropy of Astrophysical Black Holes},''
  \href{http://dx.doi.org/10.1103/PhysRevLett.132.141501}{{\em Phys. Rev.
  Lett.} {\bfseries 132} no.~14, (2024) 141501},
  \href{http://arxiv.org/abs/2212.08623}{{\ttfamily arXiv:2212.08623
  [hep-th]}}.

\bibitem{Boruch:2023trc}
J.~Boruch, L.~V. Iliesiu, and C.~Yan, ``{Constructing all BPS black hole
  microstates from the gravitational path integral},''
  \href{http://arxiv.org/abs/2307.13051}{{\ttfamily arXiv:2307.13051
  [hep-th]}}.

\bibitem{Antonini:2023hdh}
S.~Antonini, M.~Sasieta, and B.~Swingle, ``{Cosmology from random
  entanglement},'' \href{http://dx.doi.org/10.1007/JHEP11(2023)188}{{\em JHEP}
  {\bfseries 11} (2023) 188}, \href{http://arxiv.org/abs/2307.14416}{{\ttfamily
  arXiv:2307.14416 [hep-th]}}.

\bibitem{Climent:2024trz}
A.~Climent, R.~Emparan, J.~M. Magan, M.~Sasieta, and A.~Vilar~L\'opez,
  ``{Universal construction of black hole microstates},''
  \href{http://dx.doi.org/10.1103/PhysRevD.109.086024}{{\em Phys. Rev. D}
  {\bfseries 109} no.~8, (2024) 086024},
  \href{http://arxiv.org/abs/2401.08775}{{\ttfamily arXiv:2401.08775
  [hep-th]}}.

\bibitem{Iliesiu:2024cnh}
L.~V. Iliesiu, A.~Levine, H.~W. Lin, H.~Maxfield, and M.~Mezei, ``{On the
  non-perturbative bulk Hilbert space of JT gravity},''
  \href{http://arxiv.org/abs/2403.08696}{{\ttfamily arXiv:2403.08696
  [hep-th]}}.

\bibitem{Boruch:2024kvv}
J.~Boruch, L.~V. Iliesiu, G.~Lin, and C.~Yan, ``{How the Hilbert space of
  two-sided black holes factorises},''
  \href{http://arxiv.org/abs/2406.04396}{{\ttfamily arXiv:2406.04396
  [hep-th]}}.

\bibitem{Papadodimas:2015xma}
K.~Papadodimas and S.~Raju, ``{Local Operators in the Eternal Black Hole},''
  \href{http://dx.doi.org/10.1103/PhysRevLett.115.211601}{{\em Phys. Rev.
  Lett.} {\bfseries 115} no.~21, (2015) 211601},
  \href{http://arxiv.org/abs/1502.06692}{{\ttfamily arXiv:1502.06692
  [hep-th]}}.

\bibitem{Banerjee:2023liw}
S.~Banerjee, P.~Basteiro, R.~N. Das, and M.~Dorband, ``{Geometric quantum
  discord signals non-factorization},''
  \href{http://dx.doi.org/10.1007/JHEP08(2023)104}{{\em JHEP} {\bfseries 08}
  (2023) 104}, \href{http://arxiv.org/abs/2305.04952}{{\ttfamily
  arXiv:2305.04952 [hep-th]}}.

\bibitem{Blommaert:2022lbh}
A.~Blommaert, J.~Kruthoff, and S.~Yao, ``{An integrable road to a perturbative
  plateau},'' \href{http://dx.doi.org/10.1007/JHEP04(2023)048}{{\em JHEP}
  {\bfseries 04} (2023) 048}, \href{http://arxiv.org/abs/2208.13795}{{\ttfamily
  arXiv:2208.13795 [hep-th]}}.

\bibitem{Saad:2022kfe}
P.~Saad, D.~Stanford, Z.~Yang, and S.~Yao, ``{A convergent genus expansion for
  the plateau},'' \href{http://dx.doi.org/10.1007/JHEP09(2024)033}{{\em JHEP}
  {\bfseries 09} (2024) 033}, \href{http://arxiv.org/abs/2210.11565}{{\ttfamily
  arXiv:2210.11565 [hep-th]}}.

\end{thebibliography}\endgroup

\end{document}